%% file: rpo.tex
    \documentclass[final]{siamltex}

\input defs     % all definitions in defs.tex

\begin{document}
                \title{
On the state space geometry of the
Kuramoto-Sivashinsky flow in a periodic domain
                 }
                  \author{
Predrag Cvitanovi\'c\footnotemark[1],
Ruslan L. Davidchack\footnotemark[2],
    and
Evangelos Siminos\footnotemark[1]
                    }

                \maketitle

\renewcommand{\thefootnote}{\fnsymbol{footnote}}
\footnotetext[1]{
          School of Physics,
          Georgia Institute of Technology, Atlanta, GA 30332-0430, USA
                          }
\footnotetext[2]{
Department of Mathematics, University of Leicester,
            University Road, Leicester LE1 7RH, UK
                }
\renewcommand{\thefootnote}{\arabic{footnote}}

                \begin{abstract}
The continuous and discrete symmetries of the
Kuramoto-Sivashinsky system restricted to a spatially
periodic domain play a prominent role in shaping the
invariant sets of its chaotic dynamics. The continuous
spatial translation symmetry leads to relative
equilibrium (traveling wave) and relative periodic
orbit (modulated traveling wave) solutions. The
discrete symmetries lead to existence of {equilibrium} and
periodic orbit solutions, induce decomposition of state space
into invariant subspaces, and enforce certain structurally
stable heteroclinic connections between equilibria. We show,
on the example of a particular small-cell Kuramoto-Sivashinsky
system, how the geometry of its dynamical state space is
organized by a rigid `cage' built by heteroclinic connections
between equilibria, and demonstrate the preponderance of
unstable relative periodic orbits and their likely role as
the skeleton underpinning spatiotemporal turbulence in
systems with continuous symmetries. We also offer novel
visualizations of the high-dimensional Kuramoto-Sivashinsky
state space flow through projections onto low-dimensional,
PDE representation independent, dynamically invariant
intrinsic coordinate frames, as well as in terms of the
physical, symmetry invariant energy transfer rates.
                \end{abstract}



\pagestyle{myheadings}
\thispagestyle{plain}
\markboth{P.~CVITANOVI\'C, R.~L. DAVIDCHACK, AND E.~SIMINOS}
         {GEOMETRY OF THE KURAMOTO-SIVASHINSKY FLOW}

\input{intro}
\input{KSe}
\input{L22eqv}
\input{summary}

\input{ackn}

\appendix

\input{fourierRLD}
\input{lmderRLD}

\bibliography{rpo}

\end{document}

%% file: defs.tex
    \usepackage{amssymb}
    \usepackage{amsmath}
    \usepackage{amsfonts}
    \usepackage{graphicx}
    \usepackage{subfigure}
    \usepackage{dsfont}
    \usepackage{mathrsfs}
    \usepackage{color} % dvips allows for colors
    \usepackage[bookmarks,colorlinks]{hyperref}
    \hypersetup{
   pdfauthor={P. Cvitanovic, R. L. Davidchack and E. Siminos},
   pdfkeywords=Kuramoto-Sivashinsky chaos,
   pdftitle=State space geometry of chaotic Kuramoto-Sivashinsky flow
                }

\newcommand{\po}{periodic orbit}

\newcommand{\rpo}{relative periodic orbit}
\newcommand{\Rpo}{Relative periodic orbit}

\newcommand{\eqv}{equilibrium}

\newcommand{\eqva}{equilibria}
\newcommand{\Eqva}{Equilibria}
\newcommand{\reqv}{relative equilibrium}

\newcommand{\reqva}{relative equilibria}
\newcommand{\Reqva}{Relative equilibria}

\newcommand{\KS}{Kuramoto-Sivashinsky}
\newcommand{\KSe}{Kuramoto-Sivashinsky equation}
\newcommand{\pCf}{plane Couette flow}

\newcommand{\tildeL}{\ensuremath{\tilde{L}}}
\newcommand{\Lint}[1]{\frac{1}{L}\!\oint d#1\,}

\newcommand{\EQV}[1]{\ensuremath{\mathrm{E}_{#1}}}
\newcommand{\REQV}[2]{\ensuremath{\mathrm{TW}_{#1#2}}}

\newcommand{\expctE}{\ensuremath{E}}    % E space averaged

\newcommand{\rf}      [1] {~\cite{#1}}
\newcommand{\refref}  [1] {ref.~\cite{#1}}

\newcommand{\refrefs} [1] {refs.~\cite{#1}}

\newcommand{\refeq}   [1] {(\ref{#1})}

\newcommand{\reffig}  [1] {Figure~\ref{#1}}
\newcommand{\reffigs} [2] {Figures~\ref{#1} and~\ref{#2}}
\newcommand{\refFig}  [1] {Figure~\ref{#1}}

\newcommand{\reftab}  [1] {Table~\ref{#1}}
\newcommand{\refTab}  [1] {Table~\ref{#1}}

\newcommand{\refsect} [1] {sect.~\ref{#1}}

\newcommand{\refSect} [1] {Sect.~\ref{#1}}

\newcommand{\refappe} [1] {appendix~\ref{#1}}

\newcommand{\beq}{\begin{equation}}
\newcommand{\eeq}{\end{equation}}
\newcommand{\ee}[1]{\label{#1} \end{equation}}
\newcommand{\bea}{\begin{eqnarray}}

\newcommand{\continue}{\nonumber \\ }
\newcommand{\nnu}{\nonumber}
\newcommand{\eea}{\end{eqnarray}}

\newcommand{\etc}{{\em etc.}}       % etcetera in italics
\newcommand{\ie}{{that is}}     % use Latin or English?  Decide later.
\newcommand{\cf}{{\em cf.}}
\newcommand{\etal}{{\em et al.}}              % etcetera in italics

\newcommand{\jacobianM}{fundamental matrix} % standard name
\newcommand{\stabmat}{stability matrix}     % stability matrix
\newcommand{\statesp}{state space}

\newcommand{\obser}{a}      % an observable from phase space to R^n
\renewcommand\Im{\ensuremath{{\rm Im\,}}}
\renewcommand\Re{\ensuremath{{\rm Re\,}}}

\newcommand{\expct}    [1]{\left\langle {#1} \right\rangle}
\newcommand{\timeAver} [1]{\overline{#1}}
\newcommand{\pVeloc}{\ensuremath{v}}    % phase-space velocity
\newcommand{\jEigvec}[1]{\ensuremath{{\mathbf e}^{(#1)}}}   % jacobiam eigenvector
\newcommand{\ExpaEig}{\ensuremath{\Lambda}}
\newcommand{\translGen}{\ensuremath{\mathcal{L}}}

\newcommand{\eigExp}[1][ ]{\ensuremath{\lambda_{#1}}}   % complex eigenexponent
\newcommand{\eigRe}[1][ ]{\ensuremath{\mu_{#1}}}    % Re eigenexponent
\newcommand{\eigIm}[1][ ]{\ensuremath{\nu_{#1}}}    % Im eigenexponent

\newcommand\pSpace{x}       % phase space x=(q,p) coordinate
\newcommand\stagn{q}        %equilibrium/stagnation point suffix
\newcommand{\bbU}{\mathbb{U}}

\newcommand\period[1]{{T_{#1}}}         %continuous cycle period
\newcommand{\Refl}{\ensuremath{R}}
\newcommand{\Shift}{\ensuremath{\tau}}
\newcommand{\shift}{\ensuremath{\ell}}
\newcommand{\velRel}{\ensuremath{c}}    % relative state velocity
\newcommand\Lyap{\lambda}           %Lyapunov exponent

%% file: intro.tex
% intro.tex
% $Author: siminos $ $Date: 2009-10-04 22:20:49 +0300 (Sun, 04 Oct 2009) $

\section{Introduction}

Recent experimental and theoretical advances\rf{science04}
support a dynamical vision of turbulence:
For any finite  spatial resolution,
a turbulent flow follows approximately for a finite time
a pattern belonging to a
{ finite alphabet}
of admissible patterns.
The long term dynamics is
a {walk through the space of these unstable patterns}.
The question is how to characterize and classify such patterns?
Here we follow the seminal Hopf paper\rf{hopf48}, and  visualize
turbulence not as  a sequence of
spatial snapshots in turbulent evolution,
but as a trajectory in an
 infinite-dimens\-ion\-al \statesp\ in which an
instant in turbulent evolution is
a {unique} point. In the dynamical systems approach,
theory of turbulence for a given system, with given boundary conditions,
is given by
(a) the geometry of the \statesp\ and (b) the associated natural measure,
\ie,
the likelihood that asymptotic dynamics visits a given \statesp\ region.

We pursue this program in context of the \KS\ (KS) equation,
one of the simplest physically interesting spatially extended
nonlinear systems.  Holmes, Lumley and Berkooz\rf{Holmes96} offer a
delightful discussion of why this system deserves study as a staging
ground for studying turbulence in full-fledged Navier-Stokes
boundary shear flows.

Flows described by partial differential equations (PDEs) are
said to be infinite-dimens\-ion\-al because if one writes them
down as a set of ordinary differential equations (ODEs), a set
of infinitely many ODEs is needed to represent the dynamics
of one PDE. Even though their {\statesp} is thus
infinite-dimens\-ion\-al, the long-\-time dynamics of viscous
flows, such as Navier-Stokes, and PDEs modeling them, such as
Kuramoto-Sivashinsky, exhibits, when dissipation is high and
the system spatial extent small, apparent `low-dimens\-ion\-al'
dynamical behaviors. For some of these the asymptotic
dynamics is known to be confined to a finite-\-dimens\-ion\-al
{\em inertial manifold}, though the rigorous upper bounds on
this dimension are not of much use in the practice.

For large spatial extent the complexity of the spatial
motions also needs to be taken into account. The systems
whose spatial correlations decay sufficiently fast, and the
attractor dimension and number of positive Lyapunov exponents
diverges with system size are said\rf{HNZks86,man90b,cross93}
to be extensive, `spatio-temporally chaotic' or `weakly
turbulent.' Conversely, for small system sizes the accurate
description might require a large set\rf{GHCW07} of coupled
ODEs, but dynamics can still be `low-dimens\-ion\-al' in the
sense that it is characterized with one or a few positive
Lyapunov exponents. There is no wide range of scales
involved, nor decay of spatial correlations, and the system
is in this sense only `chaotic.'

For a subset of physicists and mathematicians who study
idealized `fully developed,' `homogenous' turbulence the
generally accepted usage is that the `turbulent' fluid is
characterized by a range of scales and an energy cascade
describable by statistical assumptions\rf{frisch}. What experimentalists,
engineers, geophysicists, astrophysicists actually observe
looks nothing like a `fully developed turbulence.' In the
physically driven wall-bounded shear flows, the turbulence is
dominated by unstable \emph{coherent structures}, that is,
localized recurrent vortices, rolls, streaks and like. The
statistical assumptions fail, and a dynamical systems
description from first principles is called for\rf{Holmes96}.

The set of
invariant solutions investigated here is embedded into a
finite-dimens\-ion\-al inertial manifold\rf{FNSTks85} in a
non-trivial, nonlinear way. `Geometry' in the title of this
paper refers to our attempt to systematically triangulate
this set in terms of dynamically invariant solutions (\eqva,
\po s, $\ldots$) and their unstable manifolds, in a PDE
representation and numerical simulation algorithm independent
way. The goal is to describe a given `turbulent' flow
quantitatively, not model it qualitatively by a
low-dimens\-ion\-al model. For the case investigated here, the
\statesp\ representation dimension $d \sim 10^2$ is set by
requiring that the exact invariant solutions that we compute
are accurate to $\sim 10^{-5}$.

Here comes our quandary. If we ban the words `turbulence' and
`spatiotemporal chaos' from our study of small extent
systems, the relevance of what we do to larger systems is
obscured. The exact unstable coherent structures we determine
pertain not only to the spatially small `chaotic' systems,
but also the spatially large `spatiotemporally chaotic' and
the spatially very large `turbulent' systems.
So, for the lack of more precise nomenclature, we take the
liberty of using the terms `chaos,' `spatiotemporal chaos,'
and `turbulence' interchangeably.

In previous work, the \statesp\ geometry and the natural measure for
this system have been
studied\rf{Christiansen97,LanThesis,lanCvit07} in terms of unstable
periodic solutions restricted to the antisymmetric subspace of the
KS dynamics.

The focus in this paper is on the role continuous symmetries
play in spatiotemporal dynamics. The notion of exact
periodicity in time is replaced by the notion of relative
spatiotemporal periodicity, and \reqva\ and \rpo s here play
the role the \eqva\ and \po s played in the earlier studies.
Our search for \rpo s in KS system was inspired by Vanessa
L{\'o}pez \etal\rf{lop05rel} investigation of {\rpo s} of the
Complex Ginzburg-Landau equation.  However, there is a vast
literature on {\rpo s} since their first appearance, in
Poincar\'e study of the 3-body problem\rf{ChencinerLink,rtb},
where the Lagrange points are the \reqva.  They arise in
dynamics of systems with continuous symmetries, such as
motions of rigid bodies, gravitational $N$-body problems,
molecules and nonlinear waves. Recently Viswanath\rf{Visw07b}
has found both \reqva\ and \rpo s in
the plane Couette problem.
A Hopf bifurcation of a traveling
wave\rf{AGHO288,AGHks89,Krupa90} induces a small
time-dependent modulation. Brown and Kevrekidis\rf{BrKevr96}
study bifurcation branches of \po s and \rpo s in KS system
in great detail. For our system size ($\alpha=49.04$ in their
notation) they identify a periodic orbit branch. In this
context \rpo s are referred to as `modulated traveling
waves.' For fully chaotic flows we find this notion too
narrow. We compute 60,000 \po s and \rpo s that are in no
sense small `modulations' of other solutions, hence our
preference for the well established notion of a `\rpo.'

Building upon the pioneering work of
\refrefs{KNSks90,ksgreene88,BrKevr96}, we undertake here a
study of the \KS\ dynamics for a specific system size $L =
22$, sufficiently large to exhibit many of the features
typical of `turbulent' dynamics observed in large KS systems,
but small enough to lend itself to a detailed exploration of
the  \eqva\ and \reqva, their stable/unstable manifolds,
determination of a large number of \rpo s, and a preliminary
exploration of the relation between the observed
spatiotemporal `turbulent' patterns and the \rpo s.

In presence of a continuous symmetry any solution belongs to a group
orbit of equivalent solutions. The problem: If one is to
generalize the periodic orbit theory to this setting, one needs to
understand what is meant by solutions being nearby (shadowing) when
each solution belongs to a manifold of equivalent solutions. In a
forthcoming publication\rf{SCD09b} we resolve this puzzle by implementing
symmetry reduction. Here we demonstrate that, for \rpo s visiting the
neighborhood of equilibria, if one picks any
particular solution, the universe of all other solutions is rigidly
fixed through a web of heteroclinic connections between them. This
insight garnered from study of a 1-dimens\-ion\-al \KS\ PDE is more
remarkable still when applied to the plane Couette flow\rf{GHCW07},
with 3-$d$ velocity fields and two translational symmetries.

The main results presented here are: (a) Dynamics visualized through
physical, symmetry invariant observables, such as `energy,'
dissipation rate, \etc,
and through
projections onto dynamically invariant, PDE-discretization
independent \statesp\ coordinate frames, \refsect{sec:energy}. (b)
Existence of a rigid `cage' built by heteroclinic connections
between \eqva, \refsect{sec:L22}. (c) Preponderance of
unstable \rpo s and their likely role as the skeleton underpinning
spatiotemporal turbulence in systems with continuous symmetries,
\refsect{sec:rpos}.

%% file: KSe.tex
% KSe.tex
% $Author: siminos $ $Date: 2009-10-05 23:13:22 +0300 (Mon, 05 Oct 2009) $

\section{\KSe}
\label{s-KS}

\begin{figure}[t]
\begin{center}
\includegraphics[width=0.9\textwidth]{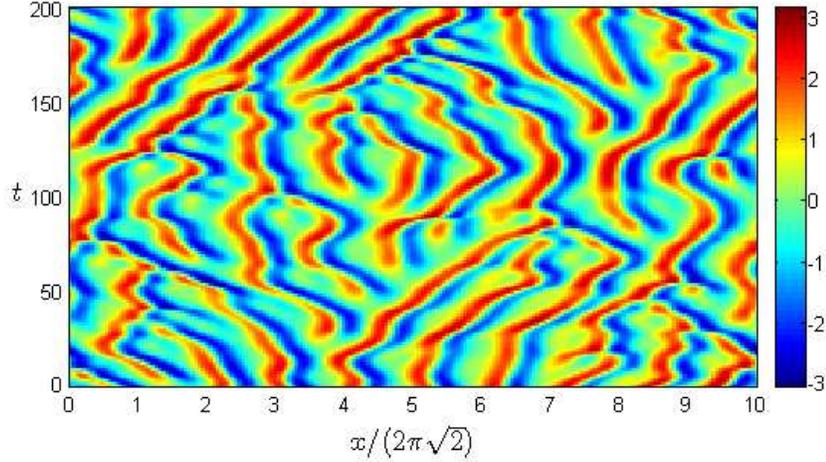} 
\end{center}
\caption{
A typical spatiotemporally chaotic solution of the \KSe, system size
$L=20\pi\sqrt{2}\approx 88.86$.  The $x$ coordinate is scaled
with the most unstable wavelength $2\pi\sqrt{2}$, which is
approximately also the mean wavelength of the turbulent flow.
The color bar indicates the color scheme for $u(x,t)$, used also
for the subsequent figures of this type.
     } \label{f:ks_largeL}
\end{figure}

The \KS\ [henceforth KS] system\rf{ku,siv},
which arises in the description of
stability of flame fronts, reaction-diffusion systems and many other
physical settings\rf{KNSks90}, is one of the simplest nonlinear PDEs that
exhibit spatiotemporally chaotic behavior. In the formulation
adopted here, the time evolution of the `flame front velocity'
$u=u(x,t)$ on a periodic domain $u(x,t) = u(x+L,t)$ is given by
\beq
  u_t = F(u) = -{\textstyle\frac{1}{2}}(u^2)_x-u_{xx}-u_{xxxx}
    \,,\qquad   x \in [-L/2,L/2]
    \,.
\ee{ks}
Here $t \geq 0$ is the time, and $x$ is the spatial coordinate.
The subscripts $x$ and $t$ denote partial derivatives with respect to
$x$ and $t$. In what follows
we shall state results of all calculations either in units of the
`dimensionless system size' $\tildeL$, or the system size $L = 2 \pi
\tildeL$. \refFig{f:ks_largeL} presents a typical `turbulent' evolution
for KS. All numerical results presented in this paper
are for the system size $\tildeL=22/2\pi = 3.5014\ldots$, for which a
structurally stable chaotic attractor is observed (see \reffig{f:ks_L22}).
Spatial periodicity $u(x,t)=u(x+L,t)$
makes it convenient to work in the Fourier space,
\beq
  u(x,t)=\sum_{k=-\infty}^{+\infty} a_k (t) e^{ i k x /\tildeL }
\,,
\ee{eq:ksexp}
with the $1$-dimensional PDE \refeq{ks}
replaced by an infinite set of
ODEs for the complex Fourier coefficients $a_k(t)$:
\beq
\dot{a}_k= \pVeloc_k(a)
     = ( q_k^2 - q_k^4 )\, a_k
    - i \frac{q_k}{2} \sum_{m=-\infty}^{+\infty} a_m a_{k-m}
\,,
\ee{expan}
where $q_k = k/\tildeL$.
Since $u(x,t)$ is real, $a_k=a_{-k}^\ast$, and we can replace the
sum by an $m > 0$ sum.

Due to the hyperviscous damping $u_{xxxx}$, long time solutions of KS
equation are smooth, $a_k$ drop off fast
with $k$, and truncations of \refeq{expan} to $16 \leq N \leq 128$
terms yield accurate solutions for system sizes considered here (see
\refappe{sec:fourierRLD}).  Robustness of the long-time dynamics
of KS as a function of the number of Fourier modes kept in truncations
of \refeq{expan} is, however, a subtle issue.  Adding an extra mode to
a truncation of the system introduces a small perturbation in the
space of dynamical systems.  However, due to the lack of structural
stability both as a function of truncation $N$, and the system size
$L$, a small variation in a system parameter can (and often will)
throw the dynamics into a different asymptotic state.  For example,
asymptotic attractor which appears to be chaotic in a $N$-dimensional
\statesp\ truncation can collapse into an attractive cycle
for $(N\!+\!1)$-dimensions.
Therefore, the selection of parameter $L$ for which a
structurally stable chaotic dynamics exists and can be
studied is rather subtle. We have found that the value of $L
= 22$ studied in \refsect{sec:L22} satisfies these
requirements. In particular, all of the equilibria and
relative equilibria persist and remain unstable when $N$ is
increased from 32 (the value we use in our numerical
investigations) to 64 and 128.  Nearly all of the \rpo s we
have found for this system also exist and remain unstable for
larger values of $N$ as well as smaller values of the
integration step size (see \refappe{sec:lmderRLD} for
details).

%%%%%%%%%%%%%%%%%%%%%%%%%%%%%%%%%%%%%%%%%%%%%%%%%%%%%%%%%%%%%%
%%%%%%%%%%%%%%%%%%%%%%%%%%%%%%%%%%%%%%%%%%%%%%%%%%%%%%%%%%%%%%%%%%

\subsection{Symmetries of \KSe}
\label{sec:KSeSymm}

The KS equation is Galilean invariant: if $u(x,t)$ is a solution,
then $u(x -ct,t) -c $, with $c$ an arbitrary constant
speed, is also a solution. Without loss of generality, in our
calculations we shall set the mean velocity of the front to zero,
\beq \int dx \, u = 0 \,. \ee{GalInv}
As $\dot{a_0}=0$ in
\refeq{expan}, $a_0$ is a conserved quantity
fixed to $a_0=0$ by the condition \refeq{GalInv}. $G$, the group of actions $ g \in G $ on a
\statesp\ (reflections, translations, \etc) is a symmetry of the KS
flow \refeq{ks} if $g\,u_t = F(g\,u)$.
The KS equation is time translationally invariant, and space translationally invariant
on a periodic domain under
the 1-parameter group of
$O(2): \{\Shift_{\shift/L},\Refl \}$.
If $u(x,t)$ is a solution, then
$\Shift_{\shift/L}\, u(x,t) = u(x+\shift,t)$
is an equivalent solution for any shift
$-L/2 < \shift \leq L/2$,
as is the
reflection (`parity' or `inversion')
\beq
    \Refl \, u(x) = -u(-x)
\,.
\ee{KSparity}
The translation operator action on the Fourier coefficients \refeq{eq:ksexp},
represented here by a complex valued vector
$a = \{a_k\in\mathbb{C}\,|\,k = 1, 2, \ldots\}$, is given by
\beq
  \Shift_{\shift/L}\, a = \mathbf{g}(\shift) \, a \,,
  \label{eq:shiftFour}
\eeq
where $\mathbf{g}(\shift) = \diag( e^{i q_k\, \shift} )$ is a complex
valued diagonal matrix, which amounts to the $k$-th mode complex plane
rotation by an angle $k\, \shift /\tildeL$.  The reflection acts on
the Fourier coefficients by complex conjugation,
\beq
  \Refl \, a = -a^\ast
\,.
\ee{FModInvSymm}
Reflection generates the dihedral subgroup $D_1 = \{1, \Refl\}$
of $O(2)$.  Let $\bbU$ be the space of
real-valued velocity fields periodic and square integrable
on the interval $\Omega = [-L/2,L/2]$,
\begin{align}
 \bbU  &= \{u \in L^2(\Omega) \; | \; u(x) = u(x+L)\}  \,.
\end{align}
A continuous symmetry maps each state $u \in \bbU$
to a manifold of functions with identical dynamic behavior.
Relation $\Refl^2 = 1$ induces linear decomposition
$u(x) = u^+(x)+ u^-(x)$,
$u^\pm(x)= P^\pm u(x) \in  \bbU^\pm$,
into irreducible subspaces
$
\bbU = \bbU^+
       \oplus \bbU^-
$, where
\beq
    P^+=(1+\Refl)/2
    \,,\qquad
    P^-=(1-\Refl)/2
\,,
\ee{P1P2proj} are the antisymmetric/symmetric projection operators.
Applying $P^+,\,P^-$ on the KS equation \refeq{ks} we have\rf{KNSks90}
\bea
 u_t^+ &=& - (u^+u^+_x + u^-u^-_x )
                - u^+_{xx} - u^+_{xxxx}
    \continue
 u_t^- &=& - (u^+u^-_x + u^-u^+_x )
                - u^-_{xx} - u^-_{xxxx}
\,.
\label{KSD1}
\eea
If $u^- = 0$, KS flow is confined to
the antisymmetric $\bbU^+$ subspace,
\beq
 u_t^+ = - u^+u^+_x
                - u^+_{xx} - u^+_{xxxx}
\,,
\label{KSU+}
\eeq
but otherwise the nonlinear terms in \refeq{KSD1}
mix the two subspaces.

Any rational shift $ \Shift_{1/m}u(x)=u(x+L/m)$ generates a discrete
cyclic subgroup $C_m$ of $O(2)$, also a symmetry of KS
system. Reflection together with $C_m$ generates another
symmetry of KS system, the dihedral subgroup $D_m$ of $O(2)$.
The only non-zero Fourier components of a solution invariant
under $C_m$ are $a_{jm} \neq 0$, $j =1,2,\cdots$, while for a
solution invariant under $D_m$ we also have the condition
$\Re a_j=0$ for all $j$.
$D_m$ reduces the dimensionality of \statesp\ and aids computation of
\eqva\ and \po s within it. For example, the 1/2-cell translations \beq
    \Shift_{1/2}\, u(x)=u(x+L/2)
\ee{KSshift}
and reflections generate $O(2)$
subgroup $D_2 = \{1, \Refl,\Shift,\Shift\Refl\}$,
which
reduces the \statesp\ into four irreducible subspaces
(for brevity, here $\Shift = \Shift_{1/2}$):
\begin{align}
 & \qquad\qquad\qquad\qquad\qquad
              ~~~ \Shift ~~ \Refl  ~\;  \Shift\Refl
    \nnu\\
P^{(1)} &= \frac{1}{4} (1 + \Shift + \Refl + \Shift\Refl)
           ~~~~  S  ~~  S   ~~   S
    \nnu\\
P^{(2)} &= \frac{1}{4} (1 + \Shift - \Refl - \Shift\Refl)
            ~~~~  S  ~~  A   ~~   A
    \nnu\\
P^{(3)} &= \frac{1}{4} (1 - \Shift + \Refl - \Shift\Refl)
           ~~~~  A  ~~  S   ~~   A
     \label{ek_defn}\\
P^{(4)} &= \frac{1}{4} (1 - \Shift - \Refl + \Shift\Refl)
          ~~~~  A  ~~  A   ~~   S
\,.
    \nnu
\end{align}
$P^{(j)}$ is the projection operator onto
$u^{(j)}$ irreducible subspace, and the last 3 columns
refer to the symmetry (or antisymmetry) of
$u^{(j)}$ functions under reflection and
1/2-cell shift.
By the same argument that identified \refeq{KSU+} as
the invariant subspace of KS, here the KS flow
stays within the
 $\bbU^S =  \bbU^{(1)}+ \bbU^{(2)}$
irreducible $D_1$ subspace of
$u$ profiles symmetric under 1/2-cell shifts.

While in general the bilinear term $(u^2)_x$  mixes the
irreducible subspaces of $D_n$, for $D_2$ there are
four subspaces invariant under the flow\rf{KNSks90}:
\begin{romannum}
 \item[$\{0\}$:~~~~~~] the $u(x)=0$ {\eqv}
 \item[$\bbU^+ = \bbU^{(1)}+ \bbU^{(3)} $:]
    the reflection $D_1$ irreducible space of antisymmetric $u(x)$
 \item[$\bbU^S =  \bbU^{(1)}+ \bbU^{(2)}$:]
    the shift $D_1$ irreducible space of $L/2$ shift symmetric  $u(x)$
 \item[$\bbU^{(1)}$:~~~~~]
    the $D_2$ irreducible  space of $u(x)$ invariant under $x\mapsto L/2-x,\ u\mapsto -u$.
\end{romannum}
With the continuous
translational symmetry eliminated within each subspace, there are no
\reqva\ and \rpo s, and one
can focus on the \eqva\ and \po s only, as was done
for $\bbU^+$ in \refrefs{Christiansen97,LanThesis,lanCvit07}.
In the Fourier
representation, the
$u \in \bbU^+$
antisymmetry amounts to having purely imaginary
coefficients, since $a_{-k}= a^\ast_k = -a_k$.
The 1/2 cell-size shift $\Shift_{1/2}$
generated 2-element discrete subgroup
$\{1,\Shift_{1/2}\}$ is
of particular interest
because in the $\bbU^+$ subspace the translational invariance of the full system reduces to
invariance under discrete translation \refeq{KSshift} by half a
spatial period $L/2$.

Each of the above dynamically invariant subspaces is unstable
under small perturbations, and generic solutions of \KSe\ belong to
the full space.
Nevertheless, since  all \eqva\ of the KS flow studied in this paper
lie in the $\bbU^+$ subspace (see
\refsect{sec:L22}), $\bbU^+$  plays important role for the global
geometry of the flow.
The linear stability matrices of these \eqva\ have
eigenvectors both in and outside of $\bbU^+$, and need to be
computed in the full \statesp.

\subsection{\Eqva\ and \reqva}
\label{sec:stks}

\Eqva\  (or the steady solutions)
are the fixed profile time-invariant solutions,
\beq
 u(x,t) = u_\stagn(x)
\,.
\ee{eqva}
Due to the translational symmetry,
the KS system also allows for
\reqva\ (traveling waves, rotating waves),
characterized by a fixed profile $u_\stagn(x)$
moving with constant speed $c$, {\ie}
\beq
 u(x,t) =  u_\stagn(x-ct)
\,.
\ee{reqva}
Here suffix ${}_\stagn$ labels a particular invariant solution.
Because of the reflection symmetry \refeq{KSparity},
the \reqva\ come in counter-traveling pairs
$u_\stagn(x-ct)$, $-u_\stagn(-x+ct)$.

The \reqv\ condition for the {\KS} PDE \refeq{ks}
is the ODE
\beq
{\textstyle\frac{1}{2}}(u^2)_x+u_{xx}+ u_{xxxx}=c \, u_x
\ee{KSeqvCond}
which can be analyzed as a dynamical system in its own right.
Integrating once we get
\beq
{\textstyle\frac{1}{2}}u^2 - c u + u_x + u_{xxx}=\expctE
\,.
\label{eq:stdks}
\eeq
This equation can be interpreted as a 3-dimen\-si\-on\-al dynamical system
with spatial coordinate $x$ playing the role of `time,'
and the integration constant \expctE\ can be interpreted as `energy,'
see \refsect{sec:energy}.

\begin{figure}[t]       \label{fig:ksBifDiag}
\begin{center}
\includegraphics[width=0.5\textwidth]{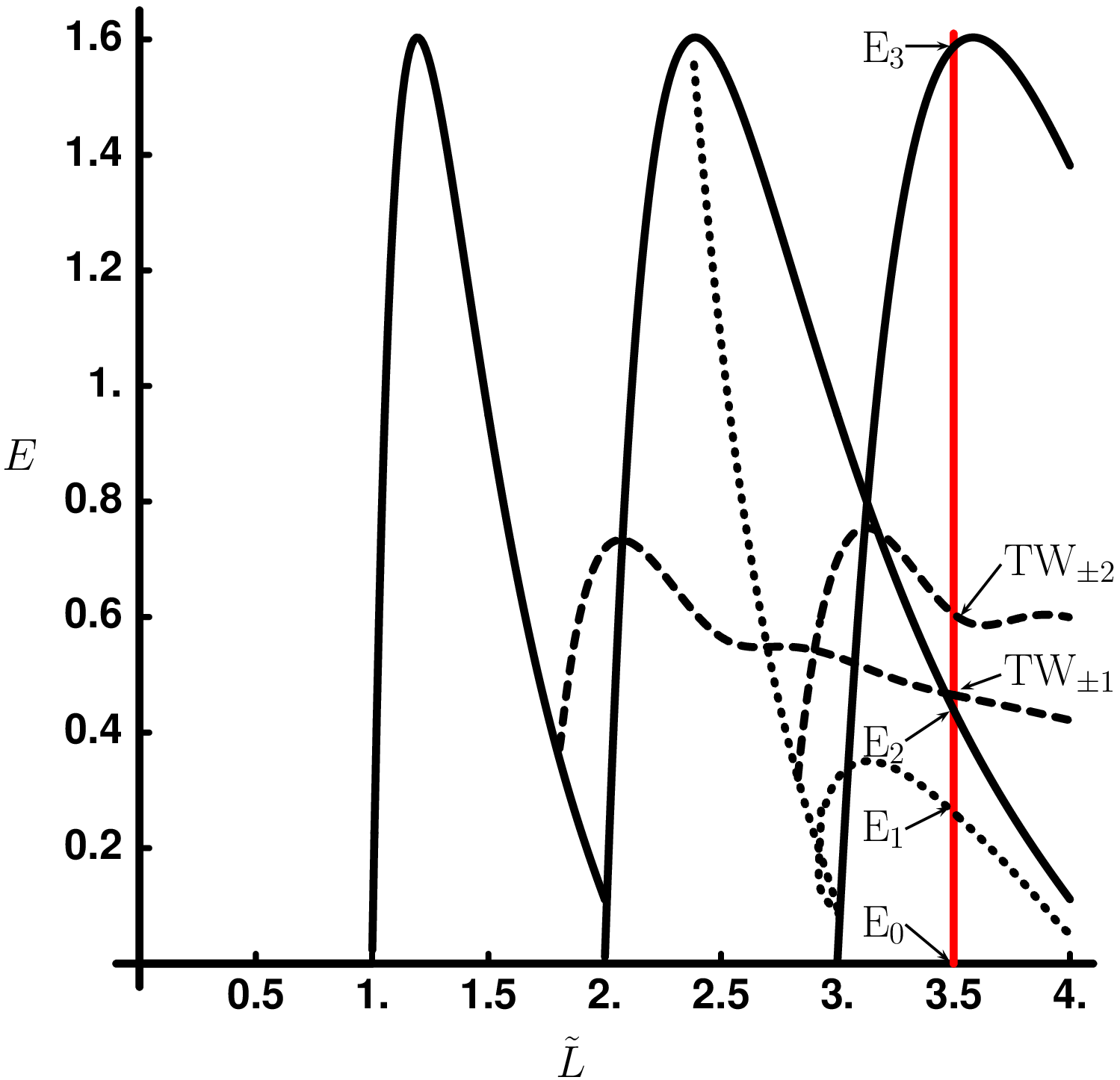}
\end{center}
\caption{
The energy \refeq{ksEnergy} of the \eqva\ and \reqva\ that
exist up to $L=22$, $\tildeL = 3.5014\ldots$, plotted as a function
of the system size $\tildeL = L/2\pi$ (additional \eqva, not present
at $L = 22$ are given in \refref{ksgreene88}). Solid curves denote
$n$-cell solutions \EQV{2} and \EQV{3}, dotted curves the GLMRT
\eqv\ \EQV{1},
and dashed curves the \reqva\ \REQV{\pm}{1} and \REQV{\pm}{2}.
The parameter $\alpha$ of \refrefs{KNSks90,ksgreene88} is
related to the system size by $\tildeL=\sqrt{\alpha/4}$.
        }
\end{figure}

For $\expctE>0$ there is rich $\expctE$-dependent dynamics,
with fractal sets of bounded solutions investigated in depth
by Michelson\rf{Mks86}. For $\tildeL<1$ the only \eqv\ of the
system is the globally attracting constant solution
$u(x,t)=0$, denoted $\EQV{0}$ from now on. With increasing
system size $L$ the system undergoes a series of
bifurcations. The resulting \eqva\ and \reqva\ are described
in the classical papers of Kevrekidis, Nicolaenko and
Scovel\rf{KNSks90}, and Greene and Kim\rf{ksgreene88},
among others. The relevant bifurcations up to the
system size investigated here are summarized in
\reffig{fig:ksBifDiag}: at $\tildeL=22/2\pi = 3.5014\cdots$,
the {\eqva} are the constant solution \EQV{0},
the  \eqv\ \EQV{1} called GLMRT by Greene and
Kim\rf{laquey74,ksgreene88},
the $2$- and $3$-cell states
\EQV{2} and \EQV{3}, and the pairs of \reqva\ \REQV{\pm}{1},
\REQV{\pm}{2}.
All \eqva\ are in the antisymmetric subspace $\bbU^+$, while
\EQV{2} is also invariant under $D_2$ and \EQV{3} under
$D_3$.

%%%%%%%%%%%%%%%%%%%%%%%%%%%%%%%%%%%%%%%%%%%%%%%%%%%%%%%%%%%%%%%%
%%%%%%%%%%%%%%%%%%%%%%%%%%%%%%%%%%%%%%%%%%%%%%%%%%%%%%%%%%%%%%%%%%

In the Fourier representation the \reqva\ time dependence is
\beq
 a_k(t) e^{-itc q_k} = a_k(0)
\,.
\ee{reqvaF}
Differentiating with respect to time, we obtain
the Fourier space version of the \reqv\ condition
\refeq{KSeqvCond},
\beq
 \pVeloc_k(a) - i q_k \velRel a_k = 0
\,,
\ee{reqvCondF}
which we solve for (time independent) $a_k$ and $c$.
Periods of spatially periodic {\eqva} are $L/n$ with integer $n$.
Every time the system size crosses  $\tildeL=n$,
$n$-cell states
are generated through pitchfork bifurcations off $u =0$
equilibrium.
Due to the translational invariance of {\KSe},
they form invariant circles
in the full \statesp.
In the $\bbU^+$ subspace considered here,
they correspond to $2n$ points, each shifted by $L/2n$.
For a sufficiently small $L$
the number of {\eqva} is small and
concentrated on the low wave-number end of the Fourier spectrum.

In a periodic box of size $L$
both \eqva\ and \reqva\ are  periodic solutions
embedded in 3-$d$ space, conveniently represented as loops in
$(u,u_x,u_{xx})$ space, see \reffig{f:KS22Equil}\,(\textit{d}).
In this representation the continuous translation symmetry
is automatic -- a rotation in the $[0,L]$ periodic domain only
moves the points along the loop. For an \eqv\ the points
are stationary in time; for \reqv\ they move in time, but in
either case, the loop remains invariant.
So we do not have the problem that we encounter in the Fourier
representation, where seen from the frame of one of the \eqva\
the rest trace out circles under the action of continuous symmetry
translations.

From \refeq{expan} we see that the origin $u(x,t) = 0$
has Fourier modes as the linear stability eigenvectors
(see \refappe{sec:stability}).  The $|k|<\tildeL$
long wavelength perturbations of the flat-front {\eqv}
are linearly unstable, while for
$|k|$ sufficiently larger than $\tildeL$ the short wavelength
perturbations are strongly contractive. The high $k$
eigenvalues, corresponding to rapid variations of the flame
front, decay so fast that the corresponding eigendirections
are physically irrelevant. Indeed, \refref{YaTaGiChRa08} shows that
the chaotic solutions of spatially extended dissipative
systems evolve within an inertial manifold spanned by a
finite number of physical modes, hyperbolically isolated from
a set of residual degrees of freedom with high $k$, themselves individually
isolated from each other.
The most unstable mode, nearest to $|k|=\tildeL/\sqrt{2}$,
sets the scale of the mean wavelength $\sqrt{2}$
of the KS `turbulent' dynamics,
see \reffig{f:ks_largeL}.

\subsection{\Rpo s, symmetries and \po s} \label{sec:KSePO}

The KS equation \refeq{ks} is time translationally invariant, and
space translationally invariant under the 1-$d$ Lie group of $O(2)$
rotations: if $u(x,t)$ is a solution, then $u(x+\shift,t)$ and
$-u(-x,t)$ are equivalent solutions for any $-L/2 < \shift \leq
L/2$.
As a result of invariance under $\Shift_{\shift/L}$,
KS equation can have \rpo\ solutions
with a profile $u_p(x)$, period $\period{p}$, and a
nonzero shift $\shift_p$
\beq
  \Shift_{\shift_p/L}u(x,\period{p}) =
  u(x+\shift_p,\period{p}) = u(x,0) = u_p(x)\,.
\label{KSrpos}
\eeq
{\Rpo s} \refeq{KSrpos} are periodic in
$\velRel_p=\shift_p/\period{p}$ co-rotating frame (see
\reffig{f:MeanVelocityFrame}), but in the stationary frame their
trajectories are quasiperiodic.  Due to the reflection symmetry
\refeq{KSparity} of KS equation, every {\rpo} $u_p(x)$ with shift
$\shift_p$ has a symmetric partner $-u_p(-x)$ with shift $-\shift_p$.

Due to invariance under reflections, KS equation can also have
\rpo s {\em with reflection}, which are
characterized by a profile $u_p(x)$ and
period $\period{p}$
\beq
  \Refl u(x+\shift,\period{p}) =
  -u(-x-\shift,\period{p}) = u(x+\shift,0) = u_p(x)
  \,,
\label{KSpos}
\eeq
giving the family of equivalent solutions
parameterized by $\shift$
(as the choice of the reflection point is arbitrary,
the shift can take any value in $-L/2 < \shift \leq L/2$).

Armbruster \etal\rf{AGHks89,AGHO288} and Brown and
Kevrekidis\rf{BrKevr96} (see also \refref{Krupa90}) link the
birth of \rpo s to an infinite period global bifurcation
involving a heteroclinic loop connecting equilibria or a
bifurcation of \reqva, and also report creation of \rpo\
branches through bifurcation of \po s.

As $\shift$ is continuous in the interval $[-L/2, L/2]$,
the likelihood of a \rpo\ with $\shift_p = 0$ shift is zero,
unless an exact periodicity is enforced by a discrete symmetry,
such as the dihedral symmetries discussed above.
If the shift $\shift_p$ of a \rpo\ with period $\period{p}$ is such
that $\shift_p /L$ is a rational number, then the orbit is
periodic with period $n\period{p}$.  The likelihood to find such \po s is
also zero.

However, due to the KS equation invariance under
the dihedral $D_n$ and cyclic $C_n$ subgroups, the following
types of \po s are possible:

{\bf (a)} The \po\ lies
within a subspace pointwise invariant under the action of
$D_n$ or $C_n$. For instance, for $D_1$ this is the
$\bbU^+$ antisymmetric subspace, $-u_p(-x) = u_p(x)$, and
$u(x,\period{p}) = u(x,0) = u_p(x)$. The periodic orbits
found in \refrefs{Christiansen97,lanCvit07} are
all in $\bbU^+$, as the dynamics is restricted to
antisymmetric subspace. For $L=22$ the dynamics in $\bbU^+$
is dominated by attracting (within the subspace)
heteroclinic connections and thus we have no periodic orbits
of this type, or in any other of the $D_n$--invariant
subspaces, see \refsect{sec:L22}.

{\bf (b)} The \po\ satisfies
\beq
	 u(x,t+\period{p})=\gamma u(x,t)\,,
	\label{eq:POspattemp}
\eeq
for some group element $\gamma\in O(2)$ such that
$\gamma^m=e$ for some integer $m$ so that the orbit repeats
after time $m \period{p}$ (see
\refref{golubitsky2002sp} for a general discussion of
conditions on the symmetry of a \po).
If an orbit is of reflection type \refeq{KSpos},
$\Refl\Shift_{\shift/L} u(x,\period{p}) =
-u(-x-\shift,\period{p}) = u(x,0)$, then it is pre-periodic
to a \po\ with period $2\period{p}$. Indeed, since
$(\Refl\Shift_{\shift/L})^2 = \Refl^2 = 1$, and the KS
solutions are time translation invariant, it follows from
\refeq{KSpos} that
\[
  u(x,2\period{p}) = \Refl\Shift_{\shift/L} u(x,\period{p}) =
  (\Refl\Shift_{\shift/L})^2 u(x,0) = u(x,0)\;.
\]
Thus any shift acquired during time $0$ to
$\period{p}$ is compensated by the opposite shift during
evolution from $\period{p}$ to $2 \period{p}$.
All periodic orbits we have found for $L=22$ are of type
\refeq{eq:POspattemp} with $\gamma=R$. Pre-periodic orbits
with $\gamma\in C_n$ have been found by Brown and
Kevrekidis\rf{BrKevr96} for KS system sizes larger than ours,
but we have not found any for $L=22$.
Pre-periodic orbits are a hallmark of any dynamical system
with a discrete symmetry, where they have a natural
interpretation as \po s in the fundamental
domain\rf{CvitaEckardt,DasBuch}.

\section{Energy transfer rates}
\label{sec:energy}

In physical settings where the observation times are much
longer than the dynamical `turnover' and Lyapunov times
(statistical mechanics, quantum physics, turbulence) periodic
orbit theory\rf{DasBuch} provides highly accurate predictions
of measurable long-time averages such as the dissipation and
the turbulent drag\rf{GHCW07}. Physical predictions have to
be independent of a particular choice of ODE representation
of the PDE under consideration and, most importantly,
invariant under all symmetries of the dynamics. In this
section we discuss a set of such physical observables for the
1-$d$ KS invariant under reflections and translations. They
offer a representation of dynamics in which the symmetries
are explicitly quotiented out. We shall use these
observables in \refsect{sec:energyL22} in order to
visualize a set of solutions on these coordinates.

The {space average} of a function $\obser = \obser(\pSpace,t) = \obser(u(x,t))$  on
the interval $L$,
\beq
    \expct{\obser} = \Lint{\pSpace}\, \obser(\pSpace,t)
    \,,
    \label{rpo:spac_ave}
\eeq
is in general time dependent.
Its mean value is given by the {time average}
\beq
\timeAver{\obser}
    =
\lim_{t\rightarrow \infty} \frac{1}{t} \int_0^t \! d\tau \, \expct{\obser}
    =
\lim_{t\rightarrow \infty} \frac{1}{t} \int_0^t \!
    \Lint{\tau}  d\pSpace\, \obser(\pSpace,\tau)
    \,.
\label{rpo:tim_ave}
\eeq
The mean value of $\obser = \obser(u_\stagn) \equiv \obser_\stagn$ evaluated on
\eqv\ or {\reqv} $u(\pSpace,t) = u_\stagn(\pSpace-ct)$, labeled by  $q$ as in 
\refeq{reqva}, is
\beq
\timeAver{\obser}_\stagn = \expct{\obser}_\stagn = \obser_\stagn\,.
\label{rpo:u-eqv} \eeq 
Evaluation of the infinite time average
\refeq{rpo:tim_ave} on a function of a \po\ or \rpo\
$u_p(\pSpace,t)=u_p(\pSpace+\shift_p,t+\period{p})$ requires only a single
$\period{p}$ traversal,
\beq
  \timeAver{\obser}_p = \frac{1}{\period{p}}
    \int_0^{\period{p}} \! d\tau \, \expct{\obser}
\,.
\label{rpo:u-cyc}
\eeq

Equation \refeq{ks} can be written as
\beq
    u_t=- V_x
        \,,\qquad
    V(x,t)={\textstyle\frac{1}{2}}u^2+u_{x} + u_{xxx}
    \,.
\ee{ksPotent}
If $u$ is `flame-front velocity' then \expctE, defined in
\refeq{eq:stdks}, can be interpreted as the mean energy
density. So, even though KS is a phenomenological
small-amplitude equation, the time-dependent $L^2$ norm
of $u$,
\beq
    \expctE=
  \Lint{\pSpace}
  V(x,t)=
  \Lint{\pSpace} \frac{u^2}{2}
  \,,
  \label{ksEnergy}
\eeq
has a physical interpretation\rf{ksgreene88} as the average `energy'
density of the flame front. This analogy to the mean kinetic energy
density for the Navier-Stokes motivates what follows.

The energy \refeq{ksEnergy} is intrinsic to the flow,
independent of the particular ODE basis set chosen to
represent the PDE. However, as the Fourier amplitudes are
eigenvectors of the translation operator, in the Fourier
space the energy is a diagonalized quadratic norm,
\beq
\expctE
          =  \sum_{k=-\infty}^{\infty} E_k
\,,\qquad
E_k =
    {\textstyle\frac{1}{2}}|a_k|^2
\,,
\ee{EFourier}
and explicitly invariant term by term
under translations
\refeq{eq:shiftFour}
and reflections \refeq{KSparity}.

Take time derivative of the energy density \refeq{ksEnergy},
substitute \refeq{ks} and integrate by parts. Total derivatives vanish
by the spatial periodicity on the $L$ domain:
\bea
   \dot{\expctE} &=&
     \expct{u_t \, u}
         = - \expct{\left({u^2}/{2} + u_{x} + u_{xxx}\right)_x u }
    \continue
    &=&
\expct{ u_x \, {u^2}/{2} + u_{x}^2 + u_x \, u_{xxx}}
    \,.
\label{rpo:ksErate}
\eea
The first term in \refeq{rpo:ksErate} vanishes by
integration by parts,
\(
3 \expct{ u_x \, u^2}= \expct{(u^3)_x} = 0
\,,
\)
and integrating the third term by parts yet again
one gets\rf{ksgreene88} that the energy variation
\beq
   \dot{\expctE} = P - D
                \,,\qquad
      P =  \expct{u_{x}^2}
                \,,\quad
      D =  \expct{u_{xx}^2}
\ee{EnRate}
balances the power $P$ pumped in by anti-diffusion $u_{xx}$
against the energy dissipation rate $D$
by hyper-viscosity $u_{xxxx}$
in the KS equation \refeq{ks}.

The time averaged energy density  $\timeAver{E}$
computed on a typical orbit goes to a constant, so
the mean values \refeq{rpo:tim_ave} of drive and dissipation
exactly balance each other:
\beq
    \timeAver{\dot{E}}  =
    \lim_{t\rightarrow \infty}
        \frac{1}{t} \int_0^t d\tau \, \dot{\expctE}
=
      \timeAver{P} - \timeAver{D}
= 0
    \,.
\ee{rpo:EtimAve}
In particular, the \eqva\
and \reqva\ fall onto the diagonal in \reffig{f:drivedrag}\,(\textit{a}),
and so do time averages computed on \po s and \rpo s:
\beq
\timeAver{E}_p =
\frac{1}{\period{p}} \int_0^\period{p}d\tau \, E(\tau)
    \,,\qquad
\timeAver{P}_p =
\frac{1}{\period{p}} \int_0^\period{p} d\tau \, P(\tau)
    =
      \timeAver{D}_p
    \,.
\label{poE}
\eeq
In the Fourier basis \refeq{EFourier} the conservation of energy on average
takes form
\beq
0 = \sum_{k=-\infty}^{\infty} ( q_k^2 - q_k^4 )\,
    \timeAver{E}_k
\,,\qquad
E_k(t) =  {\textstyle\frac{1}{2}} |a_k(t)|^2
\,.
\ee{EFourier1}
The large $k$ convergence of this series is insensitive to the
system size $L$; $\timeAver{E_k}$ have to decrease much faster than
$q_k^{-4}$.
Deviation of $E_k$ from this bound for small $k$ determines the active modes.
For \eqva\ an $L$-independent bound
    on $E$ is given by Michelson\rf{Mks86}.
The best current bound\rf{GiacoOtto05,bronski2005} on the long-time limit
of $E$
as a function of the system size $L$ scales as
$E \propto L^2$.

%% file: L22eqv.tex
% L22eqv.tex
% $Author: siminos $ $Date: 2009-10-05 23:13:22 +0300 (Mon, 05 Oct 2009) $

\section{Geometry of state space with $L=22$}
\label{sec:L22}

%%%%%%%%%%%%%%%%%%%%%%%%%%%%%%%%%%%%%%%%%%%%%%%%%%%%%%%%%%%%%%
\begin{figure}[t]
\begin{center}
\includegraphics[width=0.9\textwidth, clip=true]{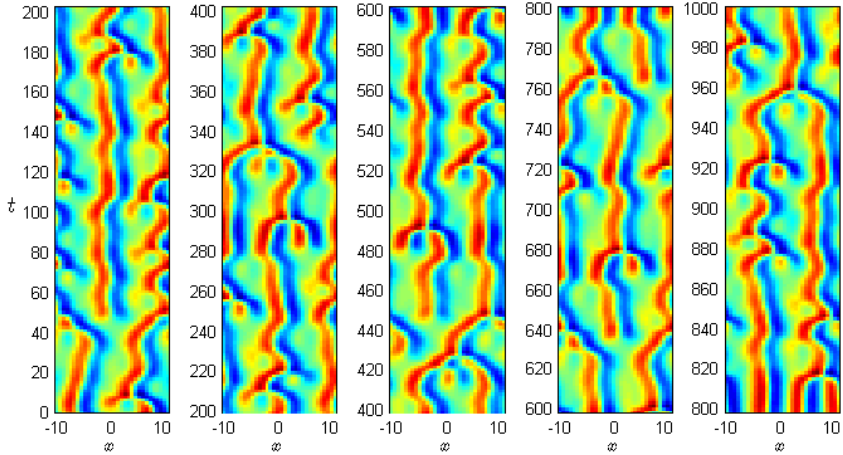}
\end{center}
\caption{
A typical chaotic orbit of the KS flow, system size $L=22$.
     } \label{f:ks_L22}
\end{figure}
%%%%%%%%%%%%%%%%%%%%%%%%%%%%%%%%%%%%%%%%%%%%%%%%%%%%%%%%%%%%%%%%%%
We now turn to exploring Hopf's vision
numerically, on a specific KS system.
An instructive example is offered by the dynamics for
the  $L=22$  system
that we specialize to for the rest of this paper.
The size of this
small system is $\sim 2.5$ mean wavelengths
($\tildeL/\sqrt{2}= 2.4758\ldots$),
and the competition between states with wavenumbers 2 and 3
leads to
what, in the context of boundary shear flows, would be
called\rf{HaKiWa95} the `empirically observed sustained
turbulence,' but in the present context may equally well be
characterized as a `chaotic attractor.' A typical long orbit
is shown in \reffig{f:ks_L22}.
Asymptotic attractor structure of small systems like
the one studied here
is very sensitive to system parameter variations, and,
as is true of
any realistic unsteady flow, there is no rigorous way of
establishing that this `turbulence' is sustained for all time,
rather than being
merely a very long transient on a way to an
attracting periodic state.
For large system size, as the one shown in \reffig{f:ks_largeL}, it is
hard to imagine a scenario under which attracting periodic states
(as shown in \refref{FSTks86}, they do exist) would have significantly
large immediate basins of attraction.
Regardless of the
(non)existence of a $t \to \infty$ chaotic attractor, study
of the invariant unstable solutions and the associated Smale
horseshoe structures in system's \statesp\ offers valuable
insights into the observed unstable `coherent structures.'

Because of the strong $k^4$ contraction, for a small system
size the long-time dynamics
is confined to low-dimensional
inertial manifold\rf{jolly_evaluating_2000}.
Indeed, numerically the covariant Lyapunov vectors\rf{ginelli-2007-99} of the
$L=22$ chaotic attractor separate into 8 ``physical''
vectors with small Lyapunov exponents
$(\Lyap_j) = (0.048,$ 0, 0, $-0.003$, $-0.189$, $-0.256$,
$-0.290$, $-0.310$),
and the remaining 54 ``hyperbolically isolated'' vectors
with rapidly decreasing exponents
$(\Lyap_j) = (-1.963$,   $-1.967$,   $-5.605$,   $-5.605$,  $-11.923$,  $-11.923$,
 $\cdots) \approx -(j/\tildeL)^4$,
in full agreement with the Yang \etal\rf{YaTaGiChRa08}
investigations of KS for large systems sizes.
%, up to $L=192$.
The chaotic dynamics mostly takes place close to a
8-dimensional manifold, with strong contraction in other
dimensions.  The two zero exponents are due to the time and
space translational symmetries of the \KSe\ and the 2 corresponding
dimensions can be quotiented out by means of discrete-time
Poincar\'e sections and $O(2)$ group orbit slices.
It was shown
in \refrefs{Christiansen97,lanCvit07} that within unstable-manifold
curvilinear coordinate frames, the dynamics on the attractor
can sometimes be well approximated by local 1- or 2-dimensional
Poincar\'e return maps.
Hence a relatively small number of real Fourier modes, such as 62
to 126 used in calculations presented here, suffices
to obtain  invariant
solutions numerically accurate to within $10^{-5}$.

We next investigate the properties of \eqva\ and \reqva\ and
determine numerically a large set of the short periods \rpo s
for KS in a periodic cell of size $L=22$.

\section{\Eqva\ and \reqva\ for $L=22$}

\begin{figure}[t]
\begin{center}
  (\textit{a})\includegraphics[width=0.35\textwidth, clip=true]{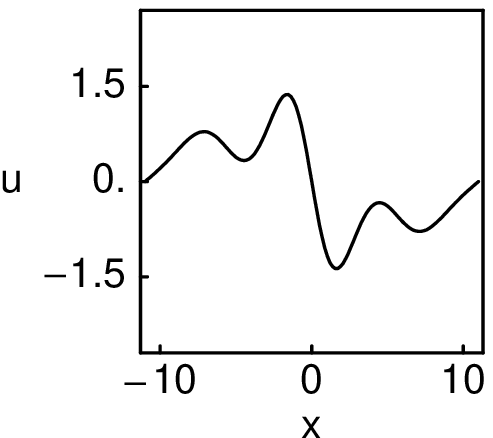}
~~~~(\textit{b})\includegraphics[width=0.35\textwidth, clip=true]{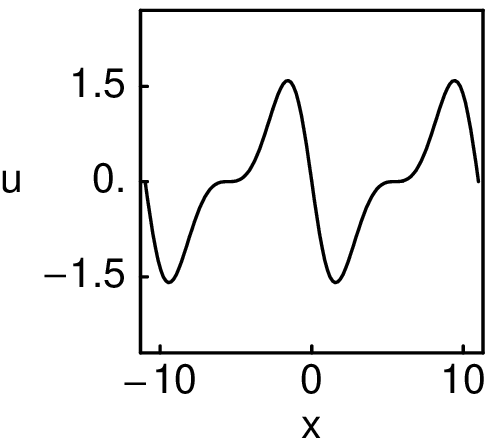}
\\
  (\textit{c})\includegraphics[width=0.35\textwidth, clip=true]{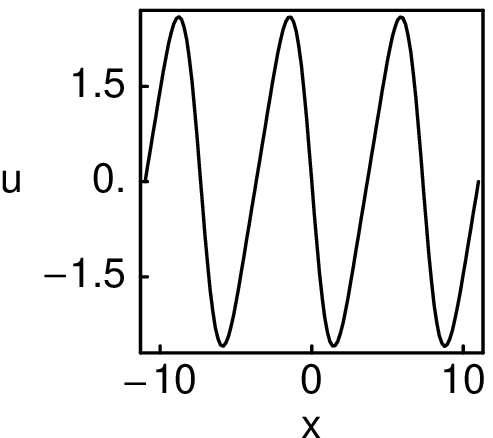}
~~~~(\textit{d})\includegraphics[width=0.35\textwidth, clip=true]{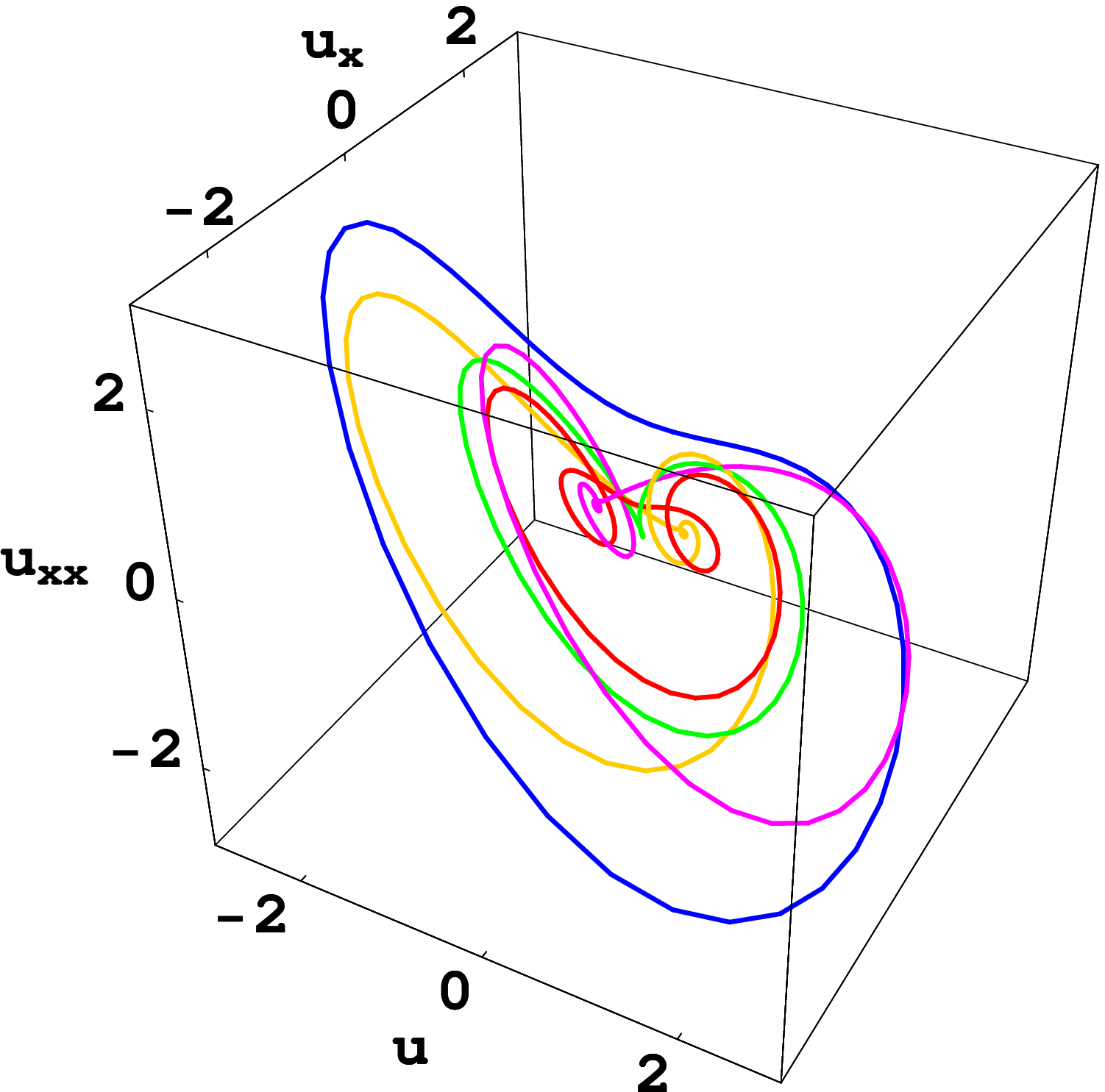}
\end{center}
\caption{
(a) \EQV{1}, (b) \EQV{2}, and (c)
\EQV{3} \eqva. The \EQV{0} \eqv\ is the $u(x)=0$ solution.
(d) $(u,u_x,u_{xx})$ representation
of (red) \EQV{1}, (green) \EQV{2},  (blue) \EQV{3} \eqva,
(purple) \REQV{+}{1},  and (orange) \REQV{-}{1} \reqva.
$L=22$ system size.
    }
\label{f:KS22Equil}
\end{figure}

In addition to the trivial \eqv\ $u=0$ (denoted \EQV{0}),
we find three \eqva\ with dominant wavenumber $k$
(denoted \EQV{k}) for $k = 1, 2, 3$.  All {\eqva}, shown in
\reffig{f:KS22Equil}, are symmetric with respect to the reflection
symmetry \refeq{KSparity}.
In addition, \EQV{2} and \EQV{3} are symmetric with respect
to translation \refeq{KSshift}, by $L/2$ and $L/3$, respectively.
\EQV{2} and \EQV{3} essentially lie in
the 2$^\mathrm{nd}$ and 3$^\mathrm{rd}$ Fourier component complex planes,
with small  deformations of the $k=2j$ and $k=3j$ harmonics, respectively.

%%%%%%%%%%%%%%%%%%%%%%%%%%%%%%%%%%%%%%%%%%%%%%%%%%%%%%%%%%%%%%%%%%
%%%%%%%%%%%%%%%%%%%%%%%%%%%%%%%%%%%%%%%%%%%%%%%%%%%%%%%%%%%%%%%%

\begin{figure}[t]
\begin{center}
\includegraphics[width=4in]{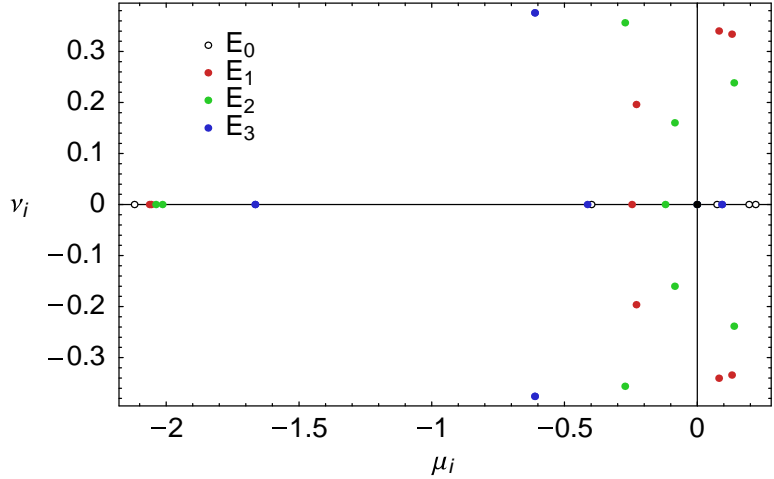}
\end{center}
\caption{
Leading  \eqv\ stability eigenvalues,
$L=22$ system size.
}
\label{f:KS22EkEigs}
\end{figure}

The stability of the {\eqva} is characterized by the eigenvalues
$\eigExp[j]$ of the \stabmat.  The leading 10 eigenvalues for each
\eqv\ are listed in \reftab{tab:Eksym};
those with $\eigRe > -2.5$ are also plotted in
\reffig{f:KS22EkEigs}.
We have computed (available upon request) the corresponding
eigenvectors as well. As an \eqv\ with $\mathrm{Re}\,
\Lyap_j > 0$ is unstable in the direction of the
corresponding eigenvector $\jEigvec{j}$, the eigenvectors
provide flow-intrinsic (PDE discretization independent)
coordinates which we use for visualization of unstable
manifolds and homo/heteroclinic connections between \eqva.
We find such coordinate frames, introduced by
Gibson \etal\rf{GHCW07,GibsonMovies}, better suited to visualization
of nontrivial solutions than the more standard Fourier mode
(eigenvectors of the $u(x,t)=0$ solution) projections.

The eigenvalues of \EQV{0} are determined by the linear part of the KS
equation \refeq{expanMvar}: $\Lyap_k=(k/\tilde{L})^2-(k/\tilde{L})^4$.
For $L=22$, there are three pairs of unstable eigenvalues, corresponding,
in decreasing order, to three unstable modes $k=2,3$, and 1.  For each
mode, the corresponding eigenvectors lie in the plane spanned by
$\Re \, a_k$ and $\Im \, a_k$. \refTab{tab:Eksym}
lists the symmetries of the stability eigenvectors of
\eqva\ \EQV{1} to \EQV{3}.

\begin{table}[t]
\caption{
Leading eigenvalues
$\eigExp[j]= \eigRe[j] \pm i\eigIm[j]$
and symmetries of the corresponding eigenvectors
of KS {\eqva} and \reqva\ for $L = 22$ system size.
We have used as our reference states the ones that lie within
the antisymmetric subspace  $\bbU^+$,
and also listed the symmetries of
the $L/4$ translated ones.
        }\label{tab:Eksym}
\begin{center} \footnotesize
\begin{tabular}{ccccc}
\EQV{1}& $\eigRe[j]$ & $\eigIm[j]$ & Symmetry & $\Shift_{1/4}\EQV{n}$ Symmetry\\\hline
  $\eigExp[1,2]$ & $\ \ 0.1308$& $0.3341$ & -  & -\\
  $\eigExp[3,4]$ & $\ \ 0.0824$& $0.3402$ & $\bbU^+$  & $\bbU^{(1)}$\\
  $\eigExp[5]$   & $0$     &          & -  & -\\
  $\eigExp[6,7]$ &$-0.2287$& $0.1963$ & $\bbU^+$  & $\bbU^{(1)}$\\
  $\eigExp[8]$   &$-0.2455$&          & -  & -\\
  $\eigExp[9]$   &$-2.0554$&          & $\bbU^+$  & $\bbU^{(1)}$\\
  $\eigExp[10]$  &$-2.0619$&          & -  & -\\[2ex]
\EQV{2}&  &  & \\\hline
  $\eigExp[1,2]$ & $\ \ 0.1390$& $0.2384$ & $\bbU^+$         & $\bbU^{(1)}$\\
  $\eigExp[3]$   & $0$      &          & $\Shift_{1/2}$        & $\Shift_{1/2}$\\
  $\eigExp[4,5]$ &$-0.0840$ & $0.1602$ & $\bbU^{(1)}$           & $\bbU^+$\\
  $\eigExp[6]$   &$-0.1194$ &          & $\Shift_{1/2}$        & $\Shift_{1/2}$\\
  $\eigExp[7,8]$ &$-0.2711$ & $0.3563$ & $\bbU^+,\,\bbU^{(1)},\,\Shift_{1/2}$  & $\bbU^+,\,\bbU^{(1)},\,\Shift_{1/2}$\\
  $\eigExp[9]$   &$-2.0130$ &          & $\bbU^{(1)}$           & $\bbU^+$\\
  $\eigExp[10]$  &$-2.0378$ &          & $\bbU^+$         & $\bbU^{(1)}$\\[2ex]
\EQV{3}&  &  & \\\hline
  $\eigExp[1]$   &$\ \ 0.0933$&          & $\bbU^+$     & $\bbU^{(1)}$\\
  $\eigExp[2]$   &$\ \ 0.0933$&          & -         & -  \\
  $\eigExp[3]$   &$0$       &          & $\Shift_{1/3}$    & $\Shift_{1/3}$\\
  $\eigExp[4]$   &$-0.4128$ &          & $\bbU^+,\,\Shift_{1/3}$  & $\bbU^{(1)},\,\Shift_{1/3}$\\
  $\eigExp[5,6]$ &$-0.6108$ & $0.3759$ & $\bbU^+$     & $\bbU^{(1)}$\\
  $\eigExp[7,8]$ &$-0.6108$ & $0.3759$ & -         & -\\
  $\eigExp[9]$   &$-1.6641$ &          & -         & -\\
  $\eigExp[10]$  &$-1.6641$ &          & $\bbU^+$     & $\bbU^{(1)}$ \\[2ex]
$\REQV{\pm}{1}$&  &  & \\\hline
  $\eigExp[1,2]$ & $\ \ 0.1156$ & $0.8173$ & -  & -\\
  $\eigExp[3,4]$ & $\ \ 0.0337$ & $0.4189$ & -  & -\\
  $\eigExp[5]$   & $0$      &          & -  & -\\
  $\eigExp[6]$   &$-0.2457$ &          & -  & -\\
  $\eigExp[7,8]$ &$-0.3213$ & $0.9813$ & -  & -\\[2ex]
$\REQV{\pm}{2}$&  &  & \\\hline
  $\eigExp[1]  $ & $\ \ 0.3370$ &          & -  & -\\
  $\eigExp[2]  $ & $0$      &          & -  & -\\
  $\eigExp[3,4]$ &$-0.0096$ & $0.6288$ & -  & -\\
  $\eigExp[5,6]$ &$-0.2619$ & $0.5591$ & -  & -\\
  $\eigExp[7,8]$ &$-0.3067$ & $0.0725$ & -  & -\\
\end{tabular}
\end{center}
\end{table}

%%%%%%%%%%%%%%%%%%%%%%%%%%%%%%%%%%%%%%%%%%%%%%%%%%%%%%%%%%%%%%%%%%
\begin{figure}[t]
\begin{center}
\includegraphics[width=0.3\textwidth, clip=true]{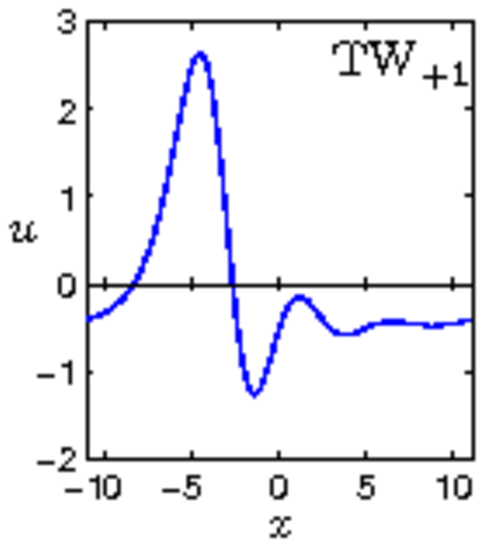}
\includegraphics[width=0.3\textwidth, clip=true]{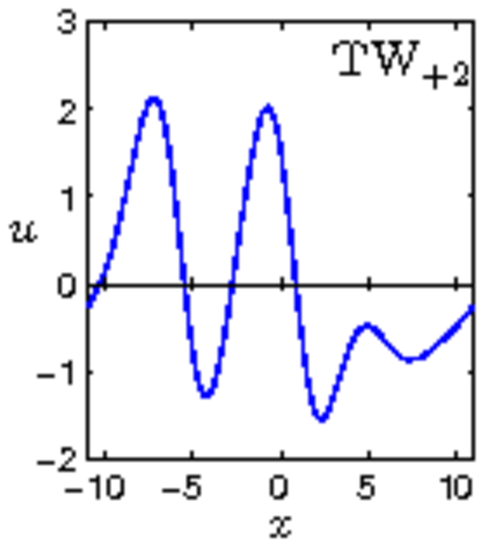}\\
\includegraphics[width=0.3\textwidth, clip=true]{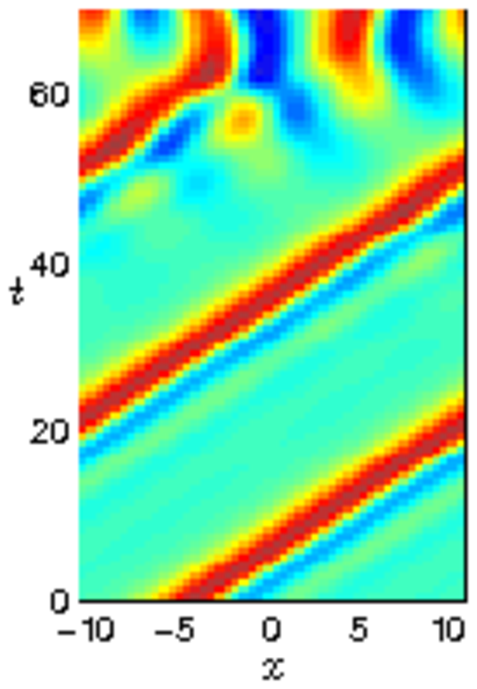}
\includegraphics[width=0.3\textwidth, clip=true]{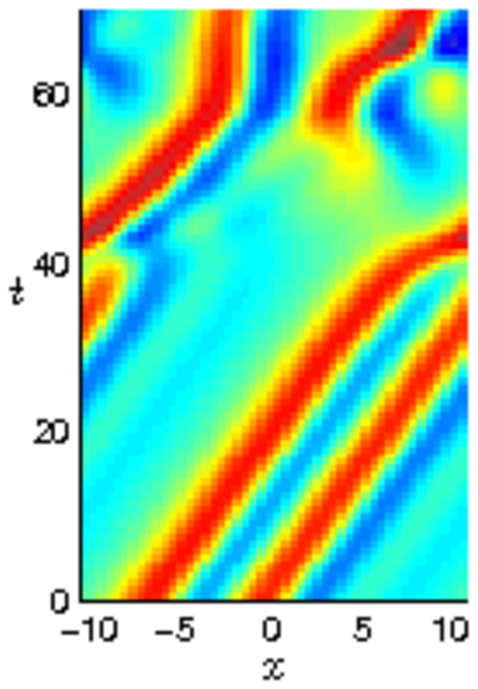}
\end{center}
\caption{
\Reqva : \REQV{+}{1} with velocity $\velRel = 0.737$ and \REQV{+}{2} with
velocity $\velRel = 0.350$.
The upper panels show the \reqva\ profiles.  The lower panels show
evolution of slightly perturbed \reqva\ and their decay into generic
turbulence. Each \reqv\ has a reflection symmetric partner related by
$u(x) \to -u(-x)$ travelling with velocity $-\velRel$.
} \label{f:ks22TW}
\end{figure}
%%%%%%%%%%%%%%%%%%%%%%%%%%%%%%%%%%%%%%%%%%%%%%%%%%%%%%%%%%%%%%%%%%

Consistent with the bifurcation diagram of \reffig{fig:ksBifDiag},
we find two pairs of \reqva\ \refeq{reqva} with velocities
$\velRel =\pm 0.73699$ and $\pm 0.34954$
which we label \REQV{\pm}{1} and \REQV{\pm}{2},
for `traveling waves.'
The profiles of the two \reqva\ and their time evolution
with eventual decay into the chaotic attractor are
shown in \reffig{f:ks22TW}.  The leading eigenvalues of
\REQV{\pm}{1} and \REQV{\pm}{2} are listed in \reftab{tab:Eksym}.

\refTab{tab:L22cminus} lists \eqv\ energy $E$,
the local Poincar\'e section return time $T$,
radially expanding Floquet multiplier $\ExpaEig_e$, and
the least contracting Floquet multiplier $\ExpaEig_c$
for all $L=22$ \eqva\ and \reqva.
The return time $T=2\pi/\eigIm[e]$ is given by the imaginary
part of the leading complex eigenvalue,
the expansion
multiplier per one turn of the most unstable spiral-out by
$\ExpaEig_e\approx\exp(\eigRe[e] T)$, and the contraction
rate along the slowest contracting stable eigendirection by
$\ExpaEig_c\approx\exp(\eigRe[c]T)$.
For \EQV{3} and \REQV{\pm}{2}, whose leading eigenvalues are
real, we use $T=1/\Lyap_1$ as the characteristic time scale.
While the complex eigenvalues set time scales of recurrences,
this time scale is useful for comparison of leading expanding
and the slowest contracting multiplier.
We learn that the shortest
`turn-over' time is $\approx 10-20$, and that if there exist
horseshoe sets of unstable \po s associated with
these \eqva,  they have unstable
multipliers of order of $\ExpaEig_e \sim 5-10$, and that
they are surprisingly thin in the folding direction, with
contracting multipliers of order of $10^{-2}$,
as also observed in \refref{lanCvit07}.

\begin{table}[ht]
    \caption{
    Properties of \eqva\ and \reqva\ determining
    the system dynamics in their vicinity.  $T$ is characteristic
    time scale of the dynamics, $\ExpaEig_e$ and $\ExpaEig_c$ are the
    leading expansion and contraction multipliers, and $E$ is the
    energy \refeq{ksEnergy}.
            }
\begin{center} \footnotesize
    \begin{tabular}{l|rrrr}
                 & $E$~~   & $T$~~  & $\ExpaEig_e$  & $\ExpaEig_c$  \\ \hline
 $\EQV{1}\ $     &\ 0.2609 &\ 18.81 &\ 11.70    &\ 0.01 \\ 
 $\EQV{2}\ $     &\ 0.4382 &\ 26.35 &\ 39.00    &\ 0.11 \\ 
 $\EQV{3}\ $     &\ 1.5876 &\ 10.72 &\ 2.72     &\ 0.01 \\ 
 $\REQV{\pm}{1}$ &\ 0.4649 &\  7.69 &\ 2.43     &\ 0.15 \\
 $\REQV{\pm}{2}$ &\ 0.6048 &\  2.97 &\ 2.72     &\ 0.97 \\ 
    \end{tabular}
\end{center}
\label{tab:L22cminus}
\end{table}

\subsection{Unstable manifolds of \eqva\ and their heteroclinic
            connections}
\label{sec:unstMnflds}

As shown in \refTab{tab:Eksym},
the \EQV{1} \eqv\ has two unstable
planes within which the solutions are spiralling out (\ie, two
pairs of complex conjugate eigenvalues).  The \EQV{2} has one such plane,
while the \EQV{3} has two real positive eigenvalues, so the solutions are
moving radially away from the \eqv\ within the plane spanned
by the corresponding eigenvectors.  Since \EQV{1} has
a larger unstable subspace, it is expected to have much less influence on the
long time dynamics compared to \EQV{2} and \EQV{3}.

\begin{figure}[t]
\begin{center}
\includegraphics[width=0.5\textwidth, clip=true]{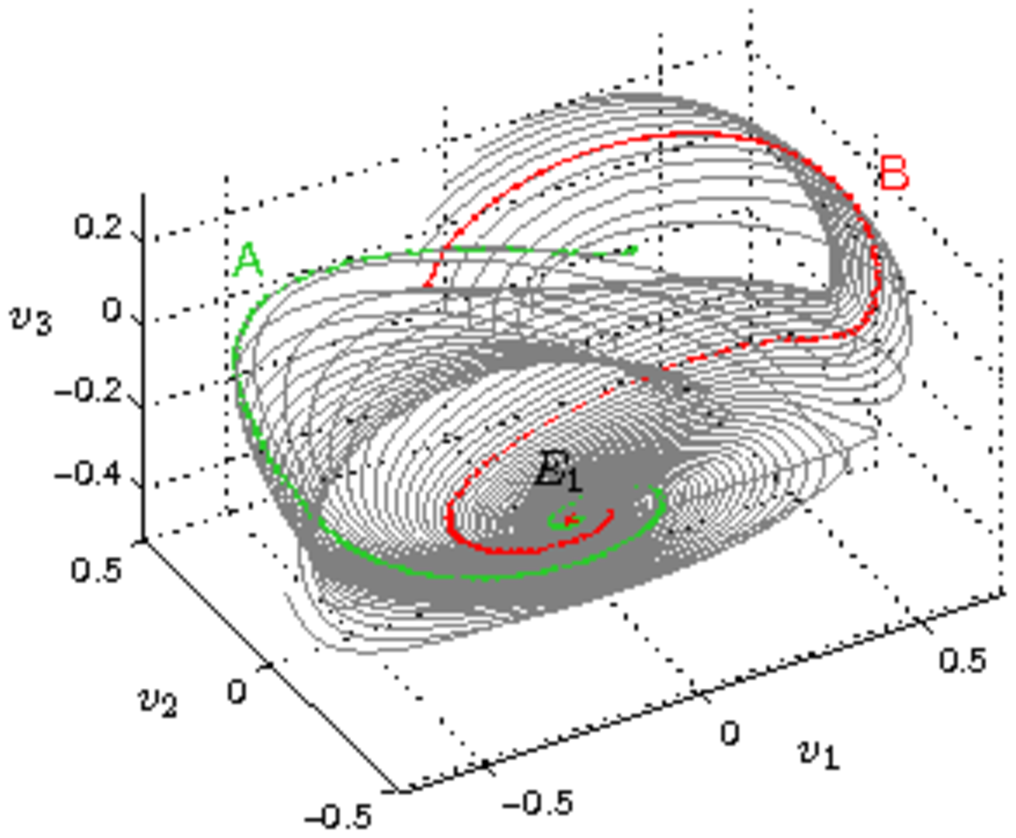}
\includegraphics[width=0.4\textwidth, clip=true]{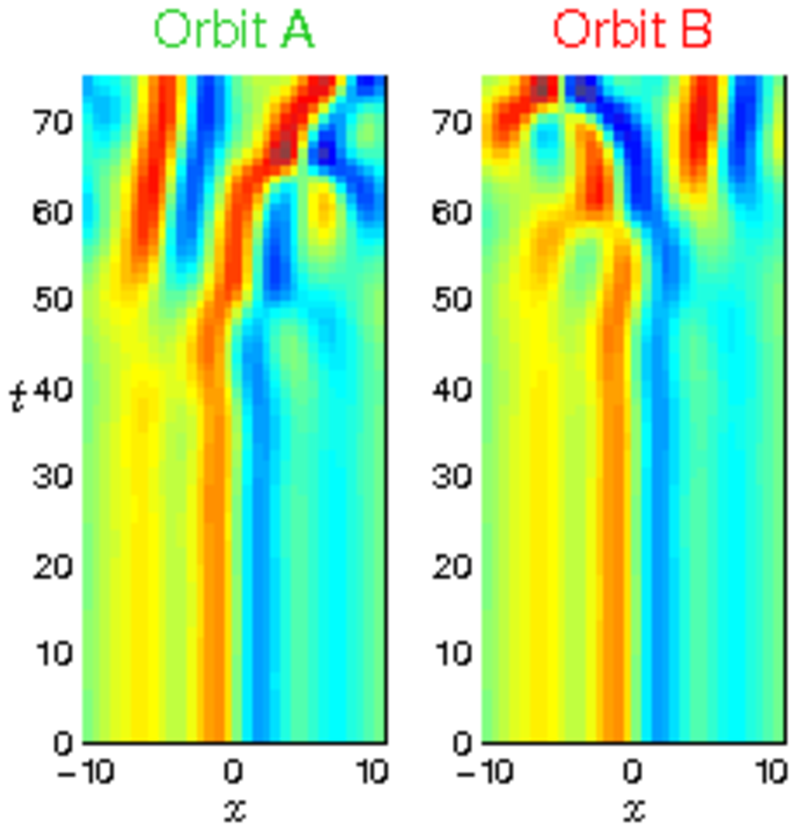}
\end{center}
\caption{
The left panel shows the unstable
manifold of \eqv\ \EQV{1} starting within the plane
corresponding to the first pair of unstable eigenvalues. The
coordinate axes $v_1$, $v_2$, and $v_3$ are
projections onto three orthonormal vectors
$\mathbf{v}_1$, $\mathbf{v}_2$, and $\mathbf{v}_3$,
respectively,
constructed from vectors
$\Re \,\jEigvec{1}$, $\Im \,\jEigvec{1}$,
and $\Re \,\jEigvec{6}$
by Gram-Schmidt orthogonalization.
The right panel shows spatial representation of two orbits $A$ and $B$.
The change of color from blue to red indicates increasing values of
$u(x)$, as in the colorbar of \reffig{f:ks_largeL}.
}
\label{f:KS22E1man1}
\end{figure}

Many methods have been developed for visualization of stable
and unstable manifolds, see \refref{krauskopf_survey_2005}
for a survey. For high-dimensional contracting flows
visualization of stable manifolds is impossible, unless the
system can be restricted to an approximate  low-dimensional
inertial manifold, as, for example, in \refref{kev01ks}. The unstable
manifold visualization also becomes harder as its dimension
increases. Here we concentrate on visualizations of $1$-- and
$2$--dimensional unstable manifolds. Our visualization is
unsophisticated compared to the methods of
\refref{krauskopf_survey_2005}, yet sufficient for our
purposes since, as we shall see, the unstable manifolds we
study terminate in another equilibrium and thus there is no
need to track them for long times.

To construct an invariant manifold containing solutions
corresponding to the pair of unstable complex conjugate eigenvalues,
$\eigExp = \eigRe \pm i\eigIm$,
$\eigRe > 0$, we start with a set of
initial conditions near \eqv\ \EQV{k},
\beq
  a(0) = a_{{\EQV{k}}} + \epsilon\,\exp(\delta)\jEigvec{j}
\,,
\ee{linUnstMan}
where $\delta$ takes a set of values uniformly distributed in the
interval $[0,2\pi\eigRe/\eigIm]$, $\jEigvec{j}$ is a unit vector in the
unstable plane, and $\epsilon > 0$ is small.

\begin{figure}[t]
\begin{center}
\includegraphics[width=0.48\textwidth, clip=true]{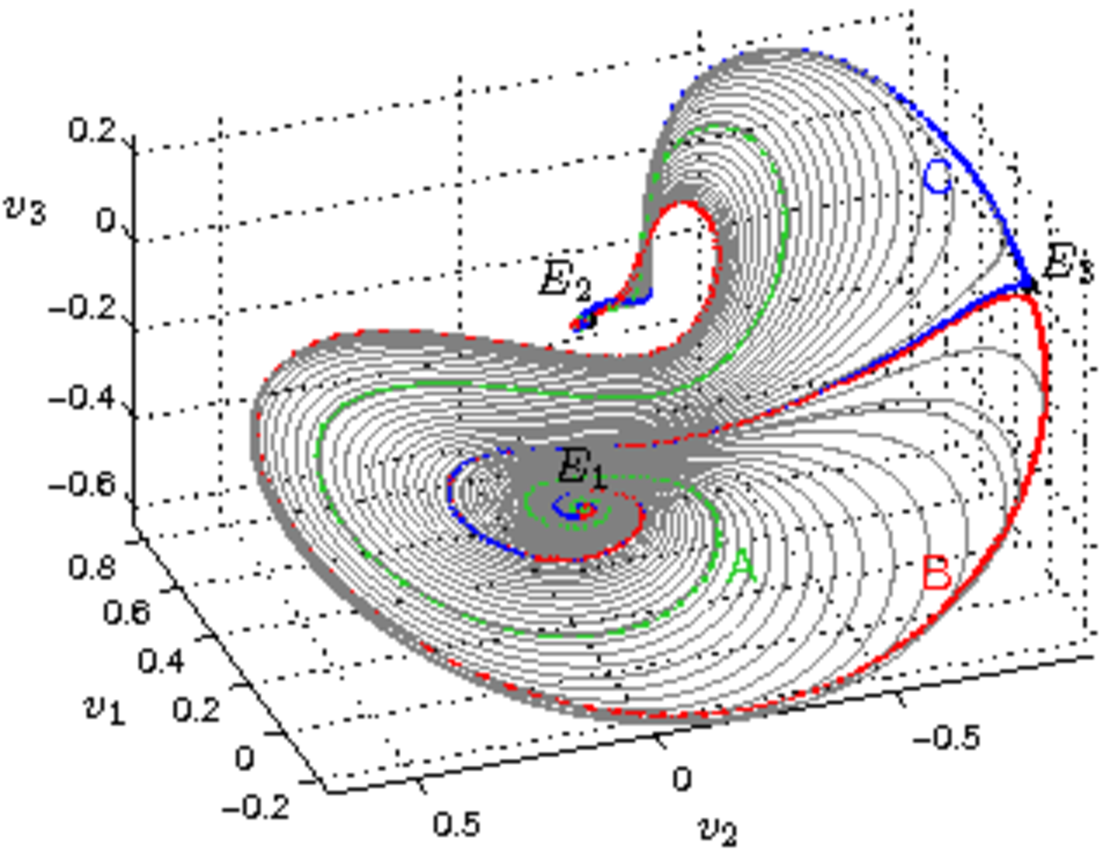}
\includegraphics[width=0.48\textwidth, clip=true]{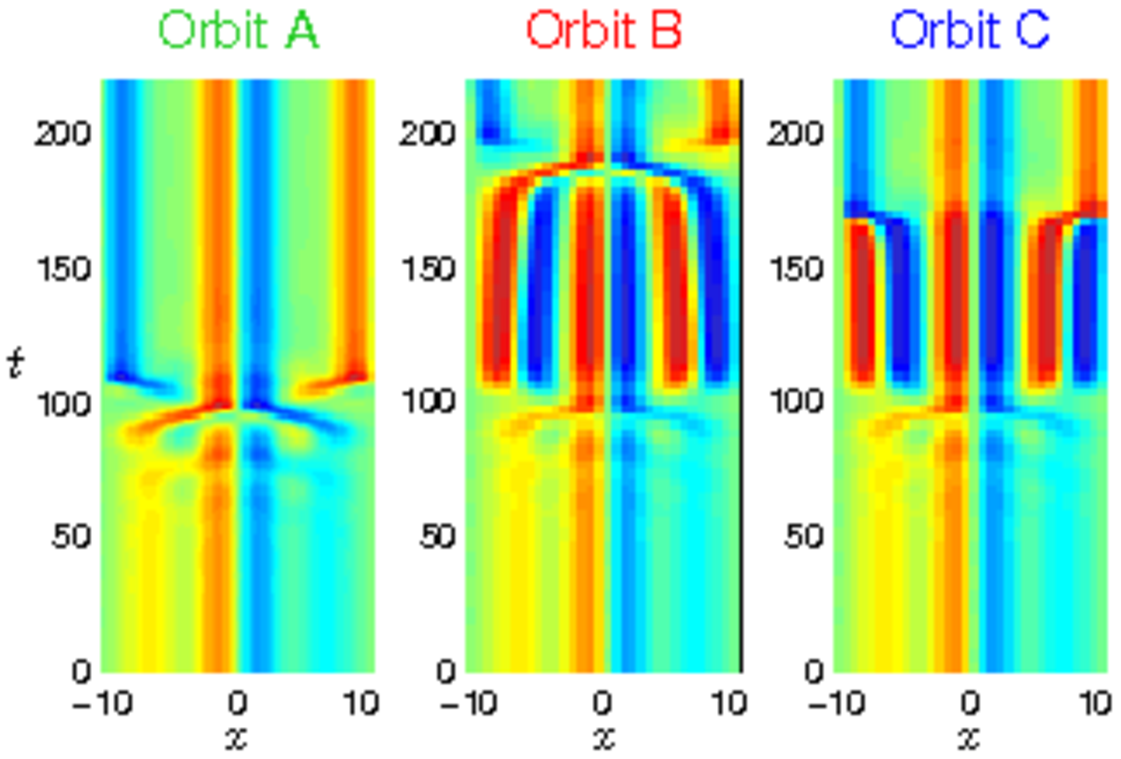}
\end{center}
\caption{
The left panel shows the unstable
manifold of \eqv\ \EQV{1} starting within the plane
corresponding to the second pair of unstable eigenvalues. The
coordinate axes $v_1$, $v_2$, and $v_3$ are
projections onto three orthonormal vectors
$\mathbf{v}_1$, $\mathbf{v}_2$, and $\mathbf{v}_3$,
respectively, constructed from vectors
\Re\, $\jEigvec{3}$, \Im\, $\jEigvec{3}$, and \Re\, $\jEigvec{6}$
by Gram-Schmidt orthogonalization.
The right panel shows spatial representation of three orbits. Orbits
$B$ and $C$ pass close to the \eqv\ \EQV{3}.
   }
\label{f:KS22E1man2}
\end{figure}

The manifold starting within the first unstable plane of \EQV{1}, with
eigenvalues $0.1308\pm i\,0.3341$, is shown in
\reffig{f:KS22E1man1}. It appears to fall directly into the
chaotic attractor.  The behavior of the manifold starting within
the second unstable plane of \EQV{1}, eigenvalues $0.0824\pm i \, 0.3402$, is
remarkably different: as can be seen in \reffig{f:KS22E1man2},
almost all orbits within the manifold converge to the \eqv\ \EQV{2}.  The
manifold also contains a heteroclinic connection from \EQV{1} to \EQV{3},
and is bordered by the $\eigExp[1]$-eigendirection
unstable manifold of \EQV{3}.

\begin{figure}[ht]
\begin{center}
\includegraphics[width=0.48\textwidth, clip=true]{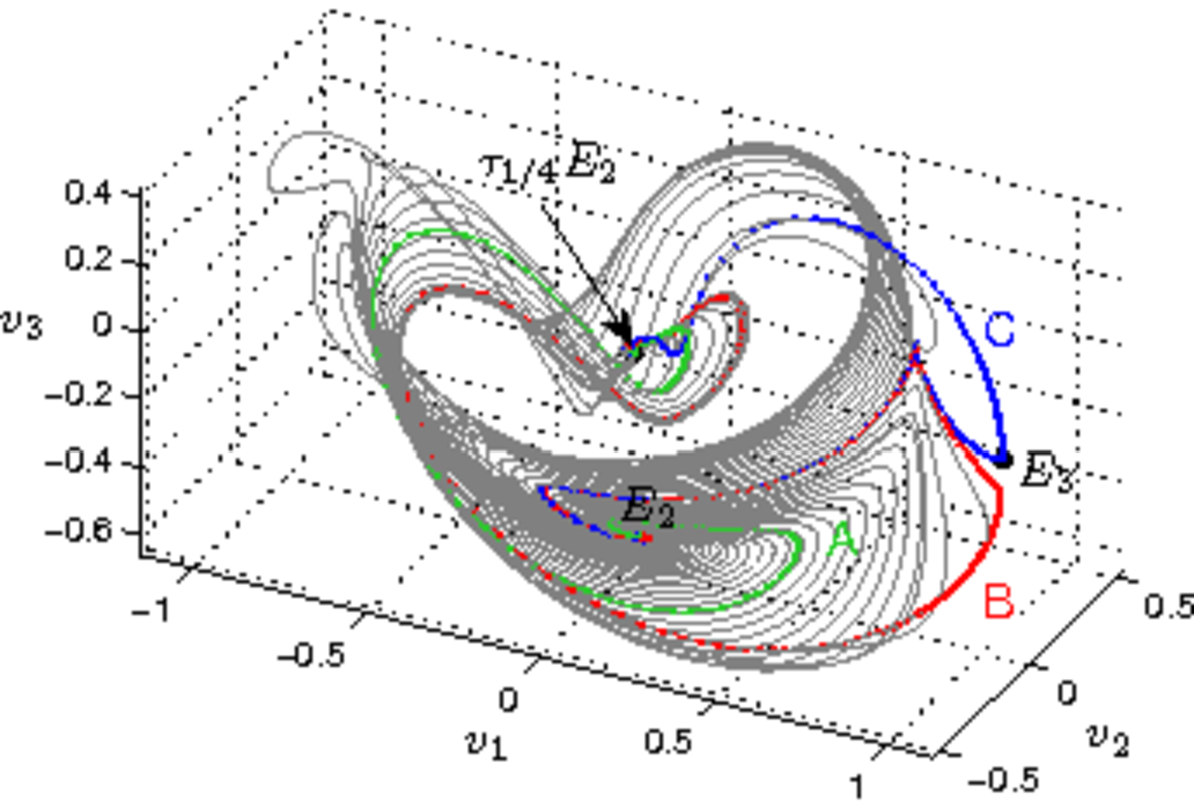}
\includegraphics[width=0.48\textwidth, clip=true]{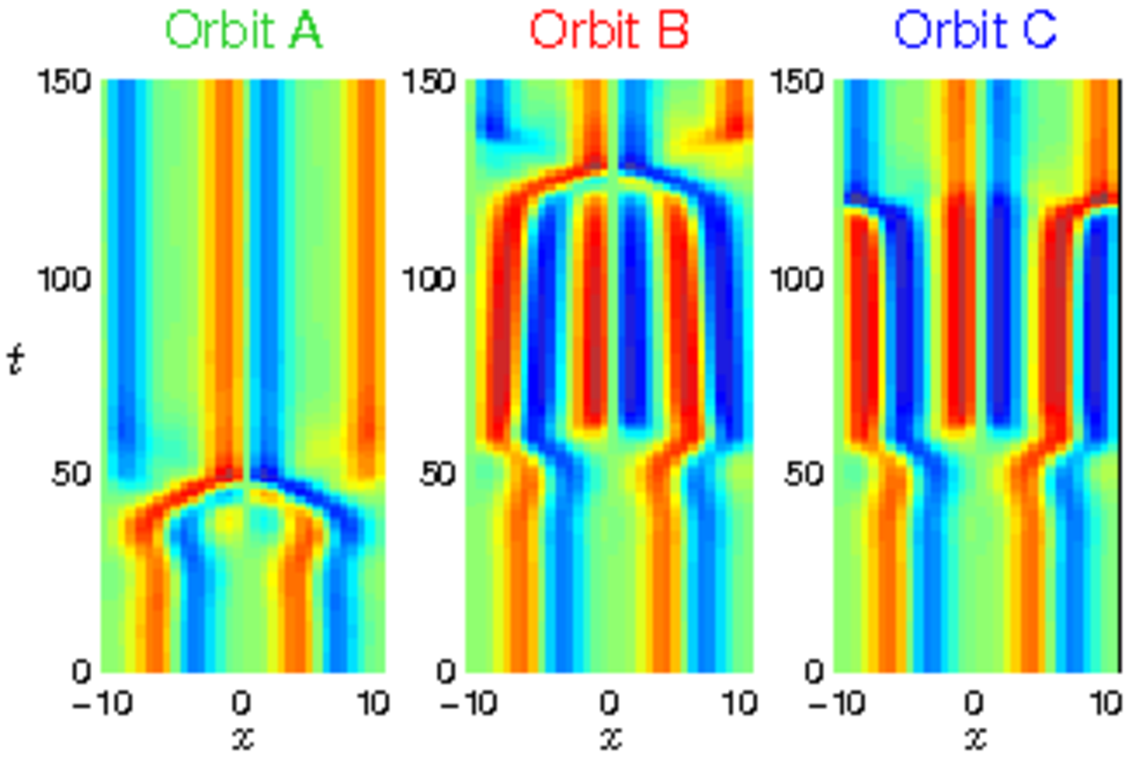}
\end{center}
\caption{
The left panel shows the two-dimensional
unstable manifold of \eqv\ \EQV{2}. The coordinate axes
$v_1$, $v_2$, and $v_3$ are
projections onto three orthonormal vectors
$\mathbf{v}_1$, $\mathbf{v}_2$, and $\mathbf{v}_3$,
respectively, constructed from vectors
\Re\, $\jEigvec{1}$, \Im\, $\jEigvec{1}$, and \Re\, $\jEigvec{7}$
by Gram-Schmidt orthogonalization.
The right panel shows spatial representation of three orbits. Orbits
$B$ and $C$ pass close to the \eqv\ \EQV{3}. See
\reffig{f:KS22Manifold} for a different visualization.
       }
\label{f:KS22E2man}
\end{figure}

%%%%%%%%%%%%%%%%%%%%%%%%%%%%%%%%%%%%%%%%%%%%%%%%%%%%%%%%%%%%%%%%
\begin{figure}[t]
\begin{center}
(\textit{a}) \includegraphics[width=0.4\textwidth, clip=true]{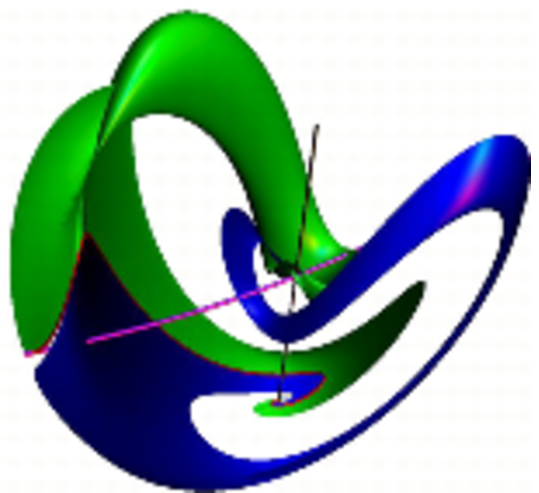}
(\textit{b}) \includegraphics[width=0.4\textwidth, clip=true]{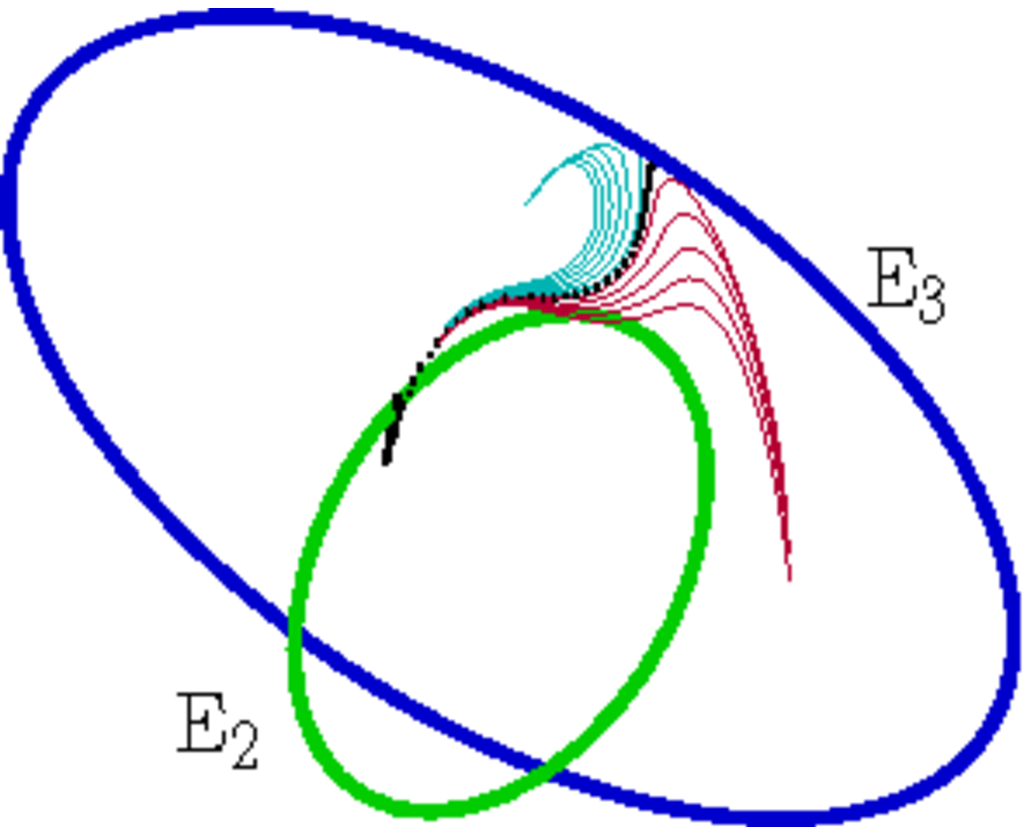}
\end{center}
\caption{
(a) (blue/green) The unstable manifold of \EQV{2}~\eqv, projection
	in the coordinate axes of \reffig{f:KS22E2man}.
    (black line) The circle of \EQV{2}~\eqva\
related by the translation invariance.
(purple line) The circle  of \EQV{3}~\eqva.
(red) The heteroclinic connection
from the \EQV{2}~\eqv\ to the \EQV{3}~\eqv\ splits
the manifold into two parts,
colored (blue) and (green).
(b) \EQV{2}~\eqv\ to \EQV{3}~\eqv\ heteroclinic
connection, $(\Re\, \jEigvec{2}, \Re\, \jEigvec{3}, ( \Im\, \jEigvec{2} + \Im\, \jEigvec{3})/\sqrt{2})$ projection. 
Here we omit the unstable manifold of \EQV{2},
keeping only a few neighboring trajectories in order to indicate
the unstable manifold of \EQV{3}. The \EQV{2} and \EQV{3}
families of \eqva\ arising from the continuous translational
symmetry of KS on a periodic domain are indicated by the two circles.
        }
\label{f:KS22Manifold}
\end{figure}
%%%%%%%%%%%%%%%%%%%%%%%%%%%%%%%%%%%%%%%%%%%%%%%%%%%%%%%%%%%%%%%%%%

The two-dimensional unstable manifold of \EQV{2} is shown in
\reffig{f:KS22E2man}.  All orbits within the manifold, except
for the heteroclinic connections from \EQV{2} to \EQV{3}, converge
to \EQV{2} shifted by $L/4$,
so this manifold, minus the heteroclinic connections, can be viewed as
a homoclinic connection.  %It also contains a pair of heteroclinic connections from
%\EQV{2} to \EQV{3}.

\begin{figure}[t]
\begin{center}
\includegraphics[width=0.45\textwidth, clip=true]{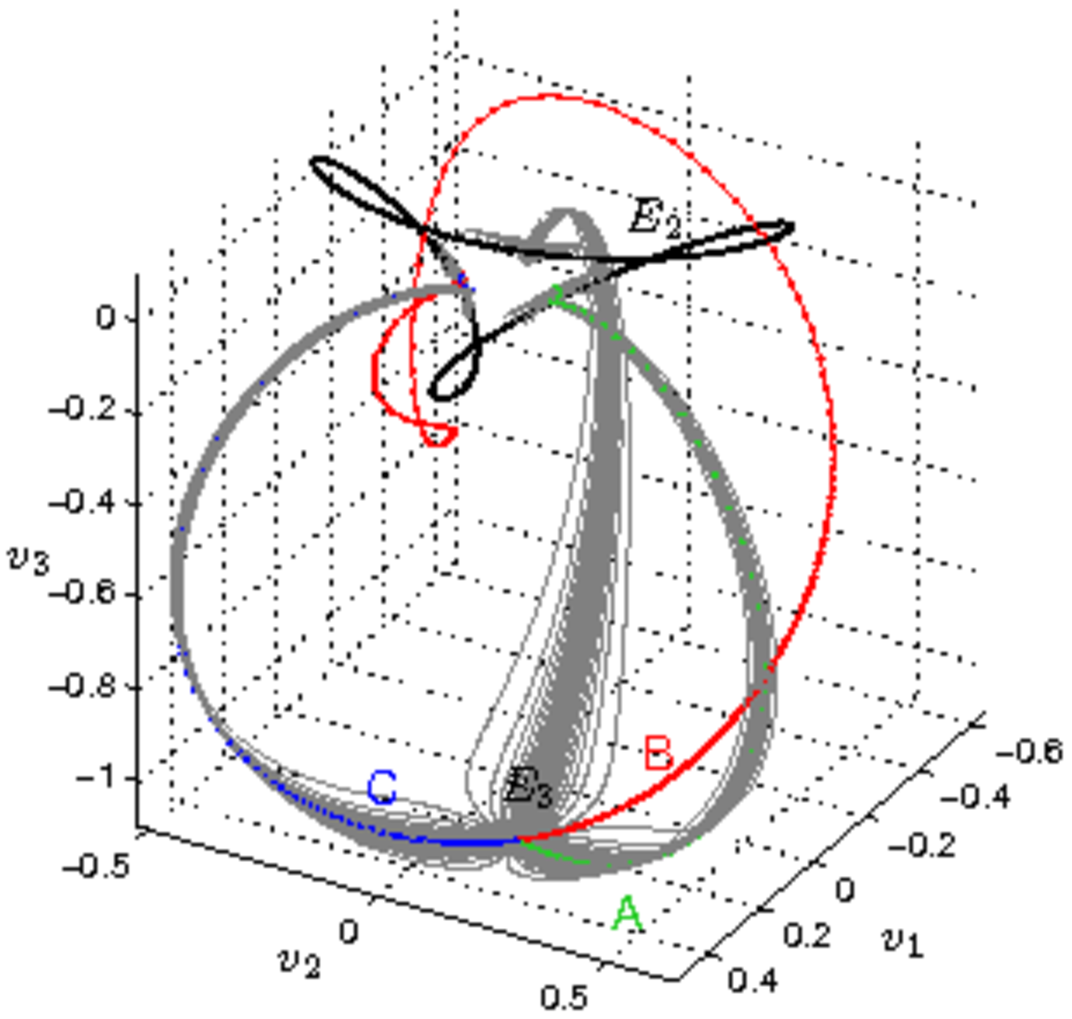}
\includegraphics[width=0.5\textwidth, clip=true]{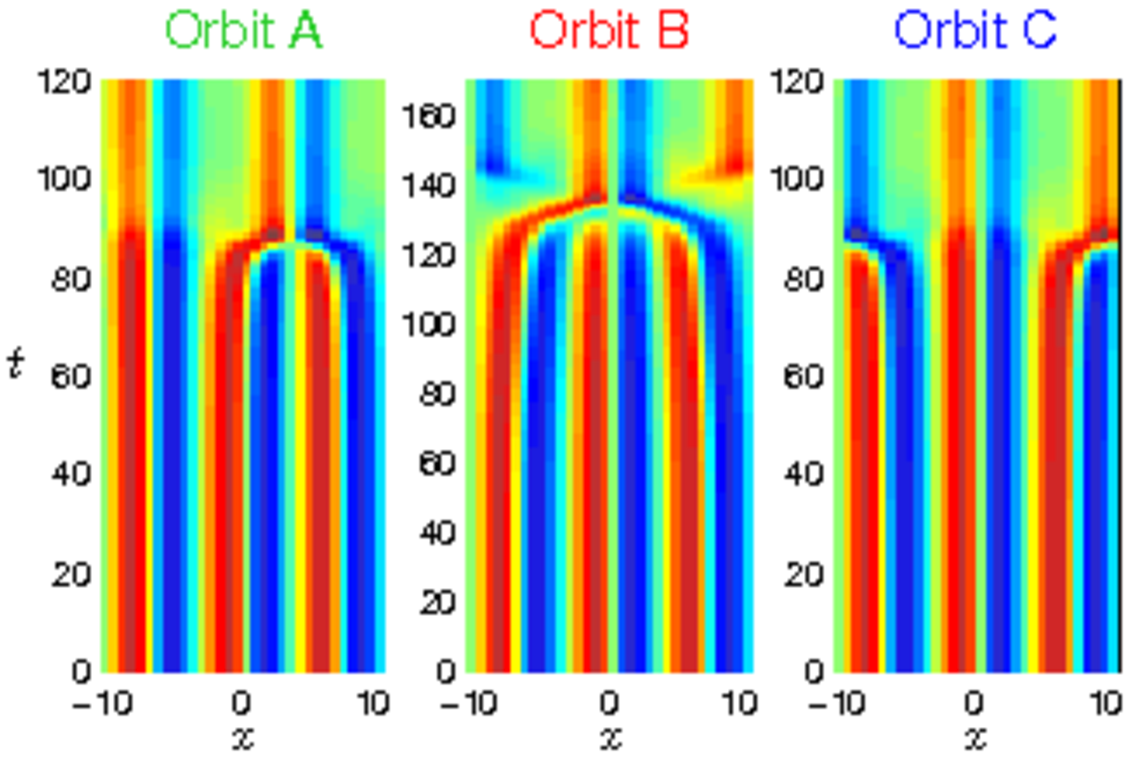}
\end{center}
\caption{
The left panel shows the two-dimensional
unstable manifold of \eqv\ \EQV{3}. The coordinate axes
$v_1$, $v_2$, and $v_3$ are
projections onto three orthonormal vectors
$\mathbf{v}_1$, $\mathbf{v}_2$, and $\mathbf{v}_3$,
respectively, constructed from vectors
$\jEigvec{1}$, $\jEigvec{2}$, and $\jEigvec{4}$ by Gram-Schmidt orthogonalization.
The black line shows a family of \EQV{2}~\eqva\ related by translational
symmetry. The right panel shows spatial representation of
three orbits. Orbits $B$ and $C$ are two different heteroclinic orbits
connecting \EQV{3} to the same point on the \EQV{2} line.
        }
\label{f:KS22E3man}
\end{figure}

The \eqv\ \EQV{3} has a pair of real unstable eigenvalues
equal to each other.  Therefore, within the plane spanned by the
corresponding eigenvectors, the orbits move radially away from
the \eqv.  In order to trace out the unstable manifold,
we start with a set of initial conditions within the unstable plane
\beq
 a(0) = a_{{\EQV{3}}} + \epsilon(\mathbf{v}_1 \cos \phi + \mathbf{v}_2 \sin \phi)\,,
  \quad\phi\in[0,2\pi]\,,
\label{unsManSeed}
\eeq
where $\mathbf{v}_1$ and $\mathbf{v}_2$ are orthonormal vectors within the
plane spanned by the two unstable eigenvectors.
%, seeded as in \refeq{linUnstMan}.
  The unstable manifold
of \EQV{3} is shown in \reffig{f:KS22E3man}.  The 3-fold symmetry of
the manifold is related to the symmetry of \EQV{3} with respect to
translation by $L/3$.  The manifold contains heteroclinic orbits
connecting \EQV{3} to three different points of the circle of {\eqva}
\EQV{2} translated set of solutions. Note also that the segments of orbits $B$ and $C$
between \EQV{3} and \EQV{2} in \reffigs{f:KS22E1man2}{f:KS22E2man}
represent the same heteroclinic connections as orbits $B$ and $C$ in
\reffig{f:KS22E3man}.

%%%%%%%%%%%%%%%%%%%%%%%%%%%%%%%%%%%%%%%%%%%%%%%%%%%%%%%%%%%%%%%%
%%%%%%%%%%%%%%%%%%%%%%%%%%%%%%%%%%%%%%%%%%%%%%%%%%%%%%%%%%%%%%%%

\begin{figure}[t]
\begin{center}
        \includegraphics[width=0.6\textwidth, clip=true]{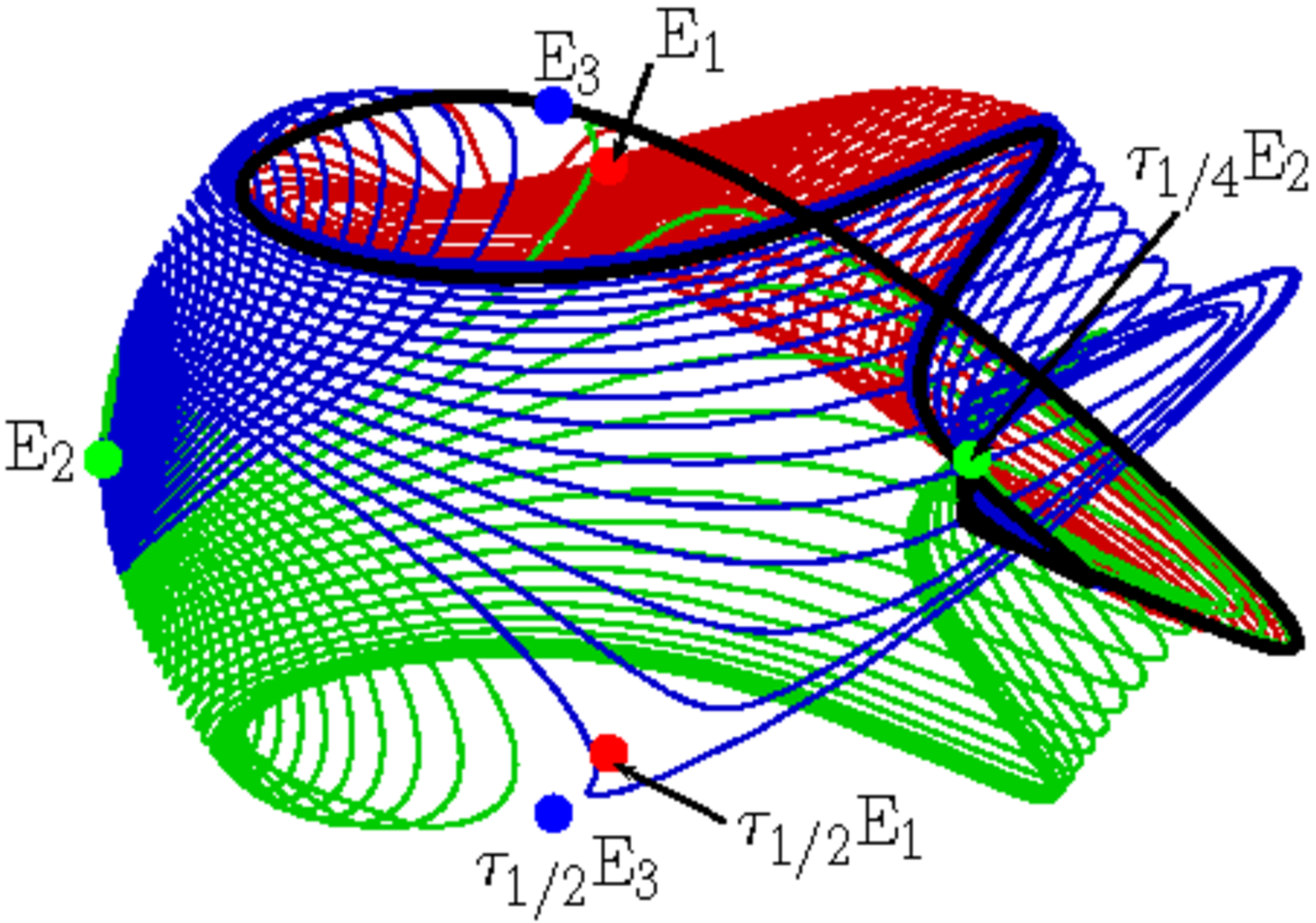}
\end{center}
\caption{ Heteroclinic connections on $\bbU^+$:
 (red) The unstable manifold of \EQV{1}~\eqv.
 (blue/green) The unstable manifold of \EQV{2}~\eqv.
 (black) Heteroclinic connections from \EQV{3}~\eqv\ to $\Shift_{1/4}$\EQV{2}~\eqv,
 where $ \Shift_{1/m}u(x)=u(x+L/m)$ is a rational shift \refeq{eq:shiftFour}.
 Projection from $128$ dimensions onto the plane given by the vectors
 $a_{\EQV{2}}-a_{\Shift_{1/4}\EQV{2}}$ and $a_{\EQV{3}}-a_{\Shift_{1/2}\EQV{3}}$.}
\label{f:KS22hetero}
\end{figure}

Heteroclinic connections are non-generic for high-dimensional
systems, but can be robust in systems with continuous
symmetry, see \refref{KrupaRobHetCyc97} for a review.
Armbruster \etal\rf{AGHks89} study a fourth order truncation
of KS dynamics on the center-unstable manifold of $\EQV{2}$
close to a bifurcation off the constant $u(x,t)=0$ solution
and prove existence of a heteroclinic connection, see also
\refref{AGHO288}. Kevrekidis \etal\rf{KNSks90} study the
dynamics numerically and establish the existence of a robust
heteroclinic connection for a range of parameters close to
the onset of the 2-cell branch in terms of the symmetry and a
flow invariant subspace. We adopt their arguments to explain
the new heteroclinic connections
shown in \reffig{f:KS22hetero} that
we have found for $L=22$.
For our system size there are exactly two representatives of
the $\EQV{2}$ family that lie in the intersection of $\bbU^+$
and $\bbU^{(1)}$ related to each other by an $L/4$ shift.
Denote them by $\EQV{2}$ and $\Shift_{1/4}\EQV{2}$
respectively. The unstable eigenplane of $\EQV{2}$ lies on
$\bbU^+$ while that of $\Shift_{1/4}\EQV{2}$ lies on
$\bbU^{(1)}$, \cf\ \reftab{tab:Eksym}. The $\EQV{3}$ family
members that live in $\bbU^+$ have one of their unstable
eigenvectors (the one related to the heteroclinic connection
to $\EQV{2}$ family)  on $\bbU^+$, while the other does not
lie on symmetry-invariant subspace. Similarly, for the
$\EQV{1}$ family we observe that the equilibria in $\bbU^+$
have an unstable plane on $\bbU^+$ (again related to the
heteroclinic connection) and a second one with no symmetry.
Thus $\Shift_{1/4}\EQV{2}$ appears as a sink on $\bbU^+$,
while all other equilibria appear as sources. This explains
the heteroclinic connections from $\EQV{1}\,,\EQV{2}$ and
$\EQV{3}$ to $\Shift_{1/4}\EQV{2}$. Observing that
$\Shift_{1/4} \bbU^+= \bbU^{(1)}$ and taking into account
\reftab{tab:Eksym} we understand that within $\bbU^{(1)}$ we
have connections from $\Shift_{1/4}\EQV{2}$ (and members of
$\EQV{1}$ and $\EQV{3}$ families) to $\EQV{2}$ and the
formation of a heteroclinic loop. Due to the translational
invariance of KS there is a heteroclinic loop for any two
points of the $\EQV{2}$ family related by an $\Shift_{1/4}$-shift.

%%%%%%%%%%%%%%%%%%%%%%%%%%%%%%%%%%%%%%%%%%%%%%%%%%%%%%%%%%%%%%%%
%%%%%%%%%%%%%%%%%%%%%%%%%%%%%%%%%%%%%%%%%%%%%%%%%%%%%%%%%%%%%%%%%%

\section{\Rpo s for $L=22$}
\label{sec:rpos}

The \rpo s satisfy the condition \refeq{KSrpos}
$u(x+\shift_p,\period{p}) = u(x,0)$,
where $\period{p}$ is the period and $\shift_p$ the phase shift.
We have limited our search to orbits with $\period{p} < 200$ and found
over 30\,000 \rpo s with $\shift_p > 0$.  The details of the algorithm
used and the search strategy employed are given in
\refappe{sec:lmderRLD}.
Each \rpo\ with phase shift
$\shift_p > 0$ has a reflection symmetric partner
$u_p(x) \to -u_p(-x)$ with phase shift $-\shift_p$.

\begin{figure}[t]
\begin{center}
\begin{tabular}{cccccc} (\textit{a}) & (\textit{c}) & (\textit{e}) &
                        (\textit{g}) & (\textit{i}) & (\textit{k})\\
\includegraphics[width=0.15\textwidth, clip=true]{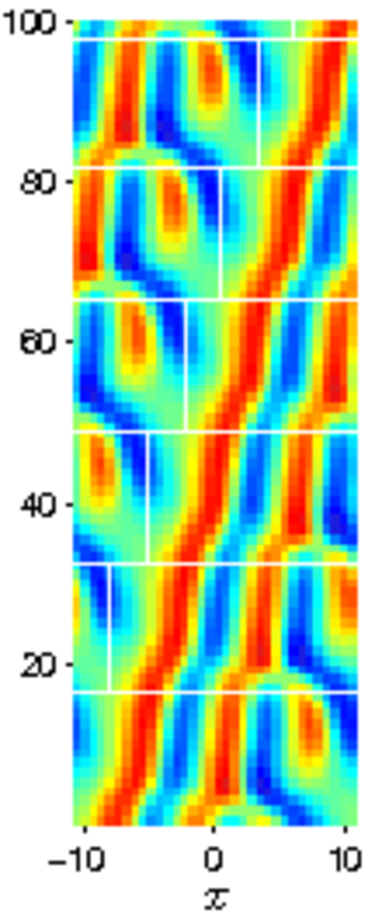}\hspace{-3ex} &
\includegraphics[width=0.15\textwidth, clip=true]{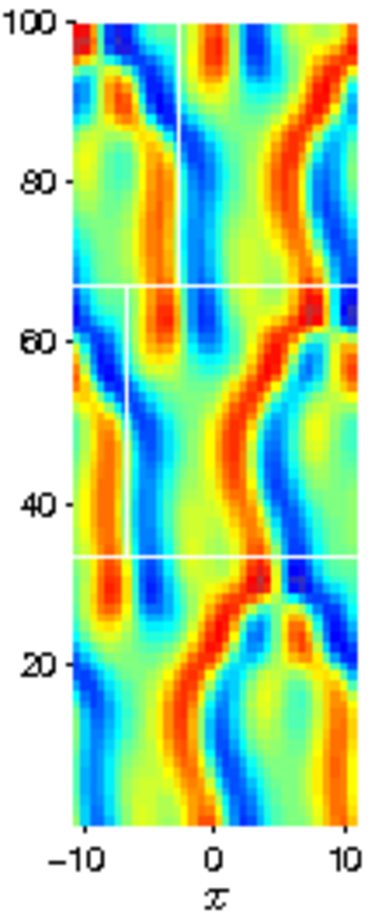}\hspace{-3ex} &
\includegraphics[width=0.15\textwidth, clip=true]{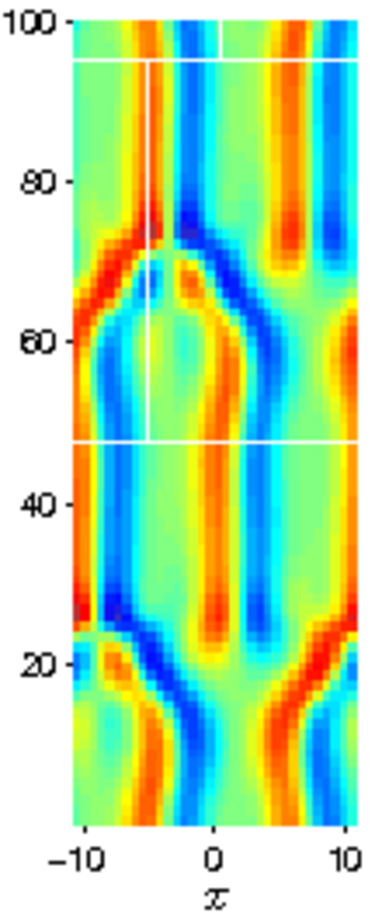}\hspace{-3ex} &
\includegraphics[width=0.15\textwidth, clip=true]{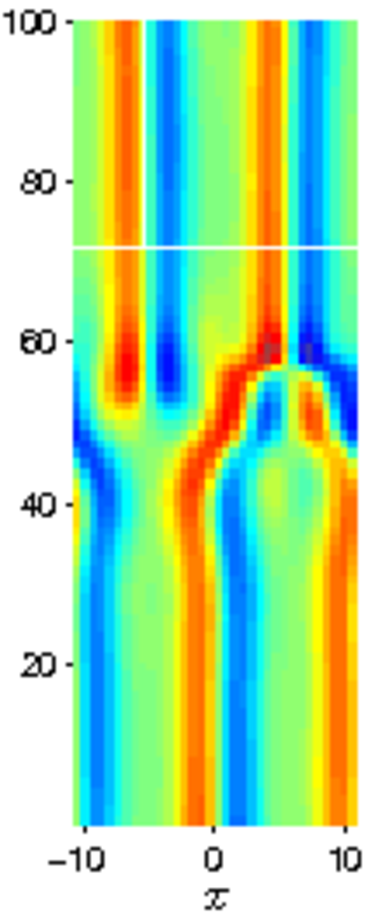}\hspace{-3ex} &
\includegraphics[width=0.15\textwidth, clip=true]{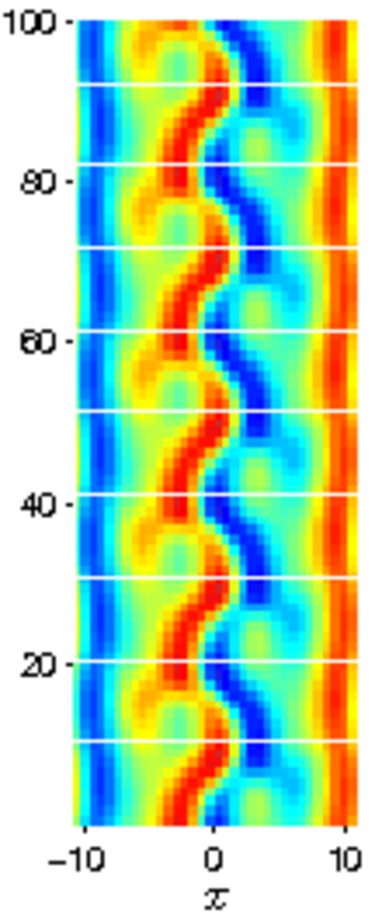}\hspace{-3ex} &
\includegraphics[width=0.15\textwidth, clip=true]{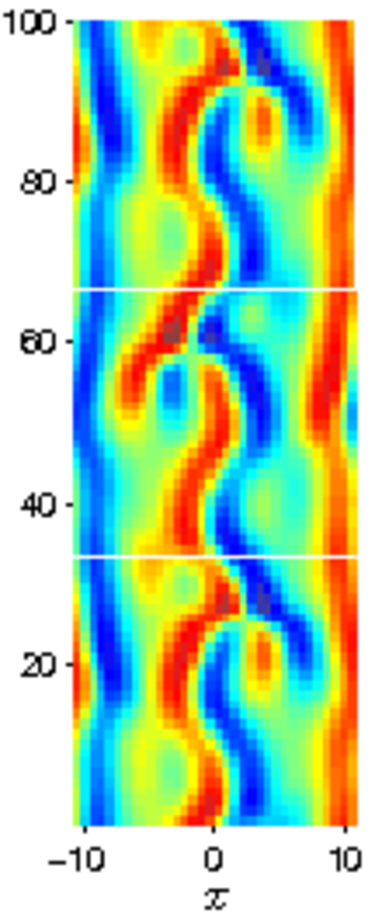}\\
(\textit{b}) & (\textit{d}) & (\textit{f}) &
(\textit{h}) & (\textit{j}) & (\textit{l})\\
\includegraphics[width=0.15\textwidth, clip=true]{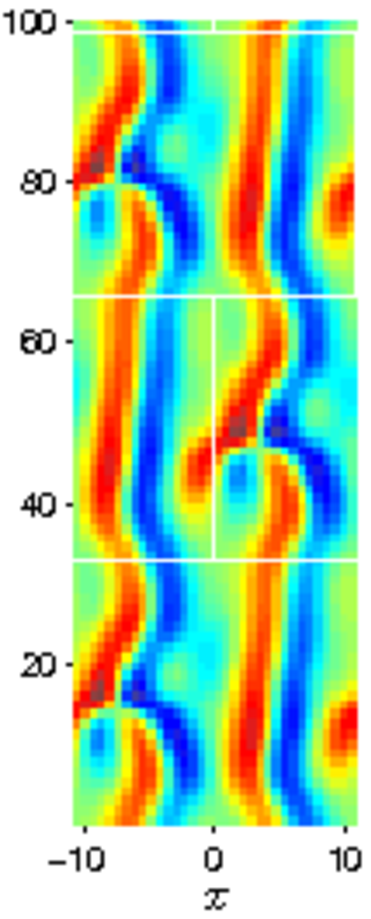}\hspace{-3ex} &
\includegraphics[width=0.15\textwidth, clip=true]{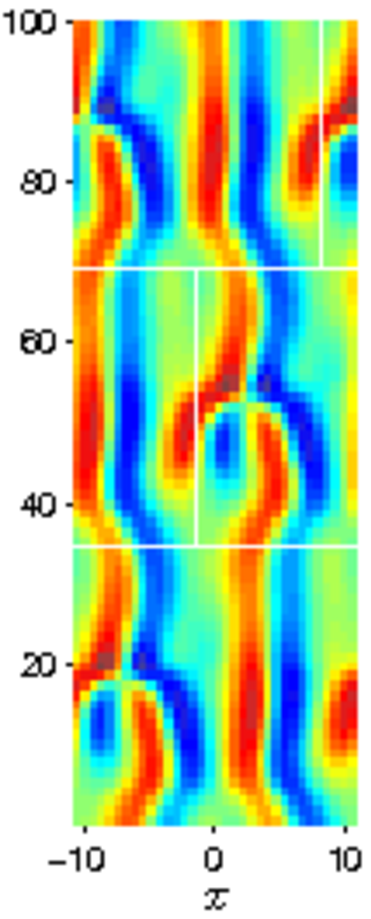}\hspace{-3ex} &
\includegraphics[width=0.15\textwidth, clip=true]{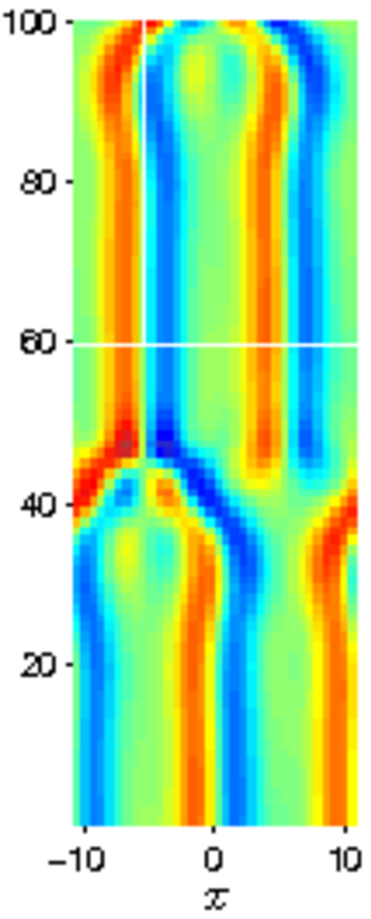}\hspace{-3ex} &
\includegraphics[width=0.15\textwidth, clip=true]{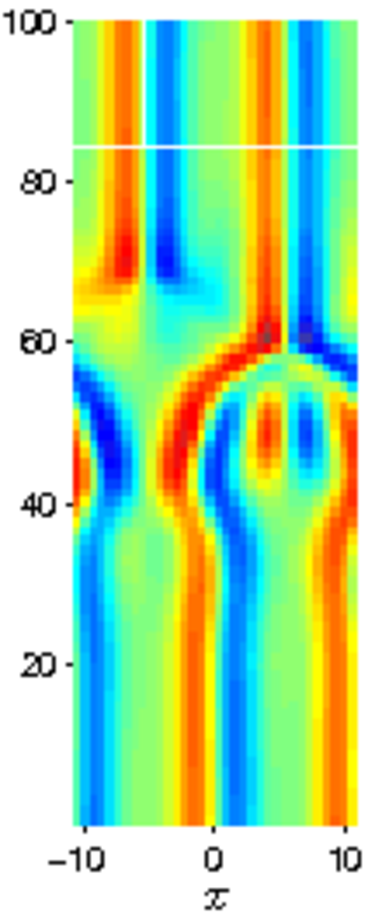}\hspace{-3ex} &
\includegraphics[width=0.15\textwidth, clip=true]{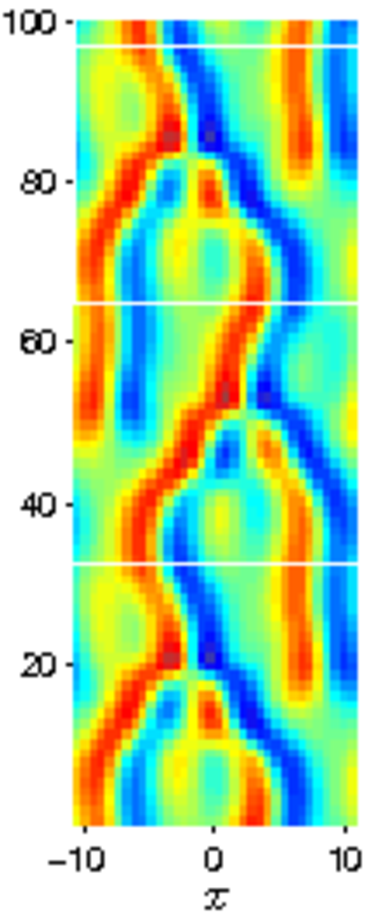}\hspace{-3ex} &
\includegraphics[width=0.15\textwidth, clip=true]{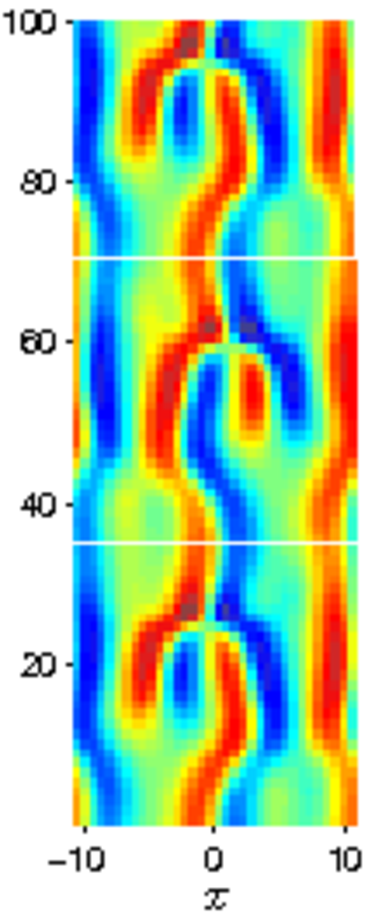}
\end{tabular}
\end{center}
\caption{Selected relative periodic and
pre-periodic
orbits of KS flow with $L = 22$:
(a) $\period{p} = 16.3$, $\shift_p = 2.86$;
(b) $\period{p} = 32.8$, $\shift_p = 10.96$;
(c) $\period{p} = 33.5$, $\shift_p = 4.04$;
(d) $\period{p} = 34.6$, $\shift_p = 9.60$;
(e) $\period{p} = 47.6$, $\shift_p = 5.68$;
(f) $\period{p} = 59.9$, $\shift_p = 5.44$;
(g) $\period{p} = 71.7$, $\shift_p = 5.503$;
(h) $\period{p} = 84.4$, $\shift_p = 5.513$;
(i) $\period{p} = 10.3$;
(j) $\period{p} = 32.4$;
(k) $\period{p} = 33.4$;
(l) $\period{p} = 35.2$.
Horizontal and vertical white lines indicate periodicity and phase
shift of the orbits, respectively.
}\label{f:ks22rpos}
\end{figure}

The small period \rpo s outline the coarse structure of the chaotic
attractor, while the longer period \rpo s resolve the finer details
of the dynamics.
The first four orbits with the shortest periods we have found are
shown in \reffig{f:ks22rpos}\,(\textit{a-d}).  The shortest \rpo\ with
$\period{p} = 16.4$ is also the most unstable, with one positive
Floquet exponent equal 0.328.  The other short orbits are less
unstable, with the largest Floquet exponent in the range
0.018 -- 0.073, typical of the long time attractor average.

%%%%%%%%%%%%%%%%%%%%%%%%%%%%%%%%%%%%%%%%%%%%%%%%%%%%%%%%%%%%%%%%
%%%%%%%%%%%%%%%%%%%%%%%%%%%%%%%%%%%%%%%%%%%%%%%%%%%%%%%%%%%%%%%%

We have found \rpo s which stay
close to the unstable manifold of \EQV{2}.
As is illustrated in \reffig{f:ks22rpos}\,(\textit{e-h}), all such orbits have
shift $\shift_p \approx L/4$, similar to the shift of orbits within
the unstable manifold of \EQV{2}, which start at \EQV{2} and
converge to $\Shift_{1/4}$\EQV{2} (see \reffig{f:KS22E2man}). This
confirms that the `cage' of unstable manifolds of equilibria plays
an important role in organizing the chaotic dynamics of the KS
equation.

\section{Pre-periodic orbits} \label{ssec:po}

As discussed in \refSect{sec:KSePO}, a \rpo\ will be
periodic, \ie, $\shift_p = 0$, if it either {\bf (a)} lives
within the $\bbU^+$ antisymmetric subspace, $-u(-x,0) =
u(x,0)$, or {\bf (b)} returns to its reflection
or its discrete rotation after a period:
$u(x,t+\period{p})=\gamma u(x,t)$, $\gamma^m=e$,
and is thus periodic with period $m\period{p}$.
The dynamics of KS flow in the antisymmetric subspace and \po
s with symmetry {\bf (a)} have been investigated
previously\rf{Christiansen97,LanThesis,lanCvit07}.
The KS flow does not have any periodic orbits of this type
for $L = 22$.

Using the algorithm and strategy described in
\refappe{sec:lmderRLD}, we have found over 30\,000
pre-periodic orbits with $\period{p} < 200$ which possess the
symmetry of type {\bf (b)} with $\gamma=\Refl\in D_1$.
Some of the shortest such orbits we have found are shown in
\reffig{f:ks22rpos}\,(\textit{i-l}). Several were found as
repeats of pre-periodic orbits during searches for \rpo s
with non-zero shifts, while most have been found as solutions
of the pre-periodic orbit condition \refeq{KSpos} with
reflection, which takes form
\beq
 -\mathbf{g}(-\shift)a^\ast(\period{p}) = a(0)\,.
\label{KSposFour}
\eeq
in the Fourier space representation
(compare this to the condition \refeq{eq:system} for \rpo s).

\section{Energy transfer rates  for $L=22$}
\label{sec:energyL22}

%%%%%%%%%%%%%%%%%%%%%%%%%%%%%%%%%%%%%%%%%%%%%%%%%%%%%%%%%%%%%%%%
\begin{figure}[t]
\begin{center}
 \begin{tabular}{cc}
        ~~~~~~~~(\textit{a})                        &   ~~~~~~~~(\textit{b}) \\
    \includegraphics[width=0.46\textwidth, clip=true]{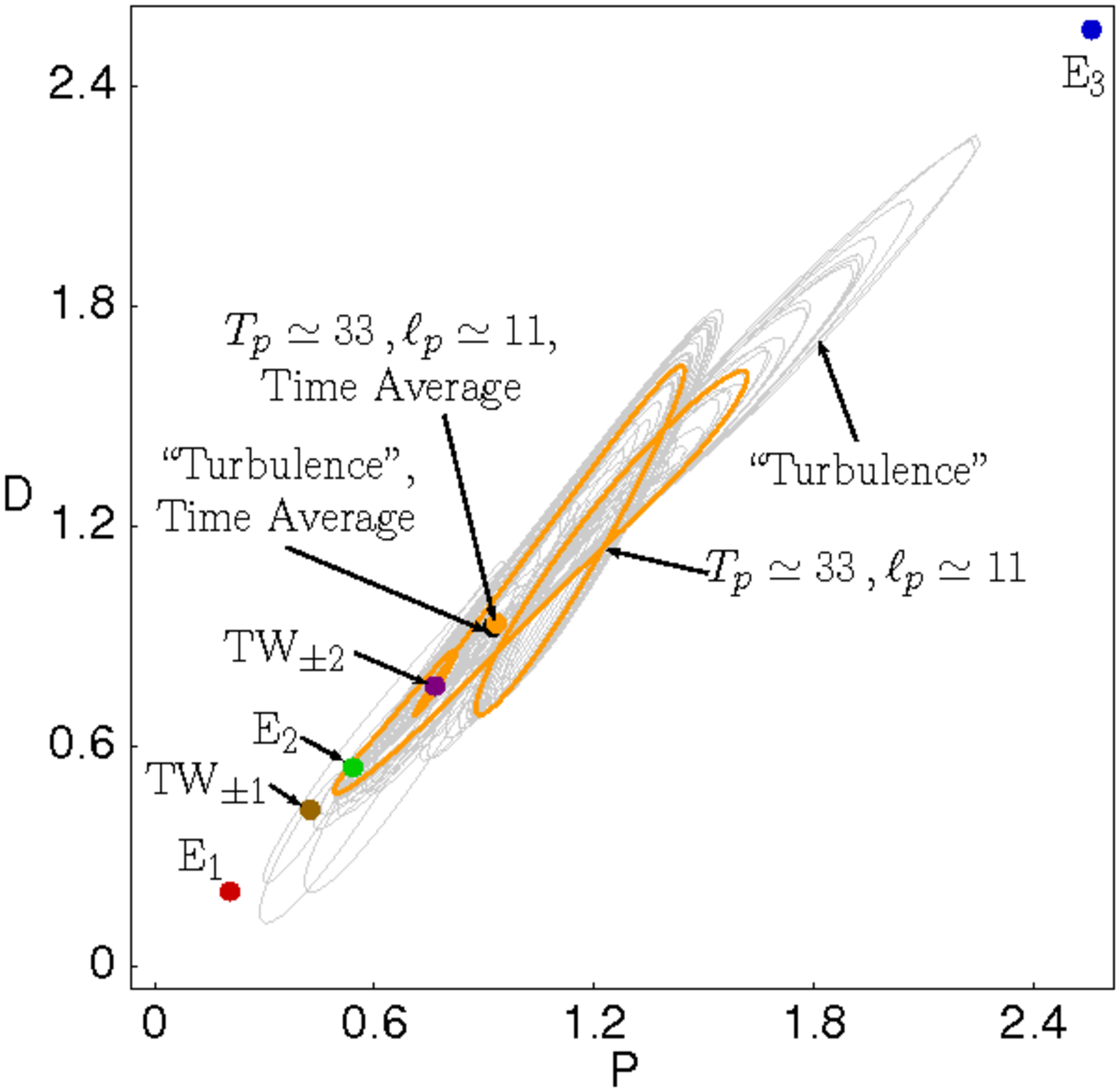}  & \includegraphics[width=0.46\textwidth, clip=true]{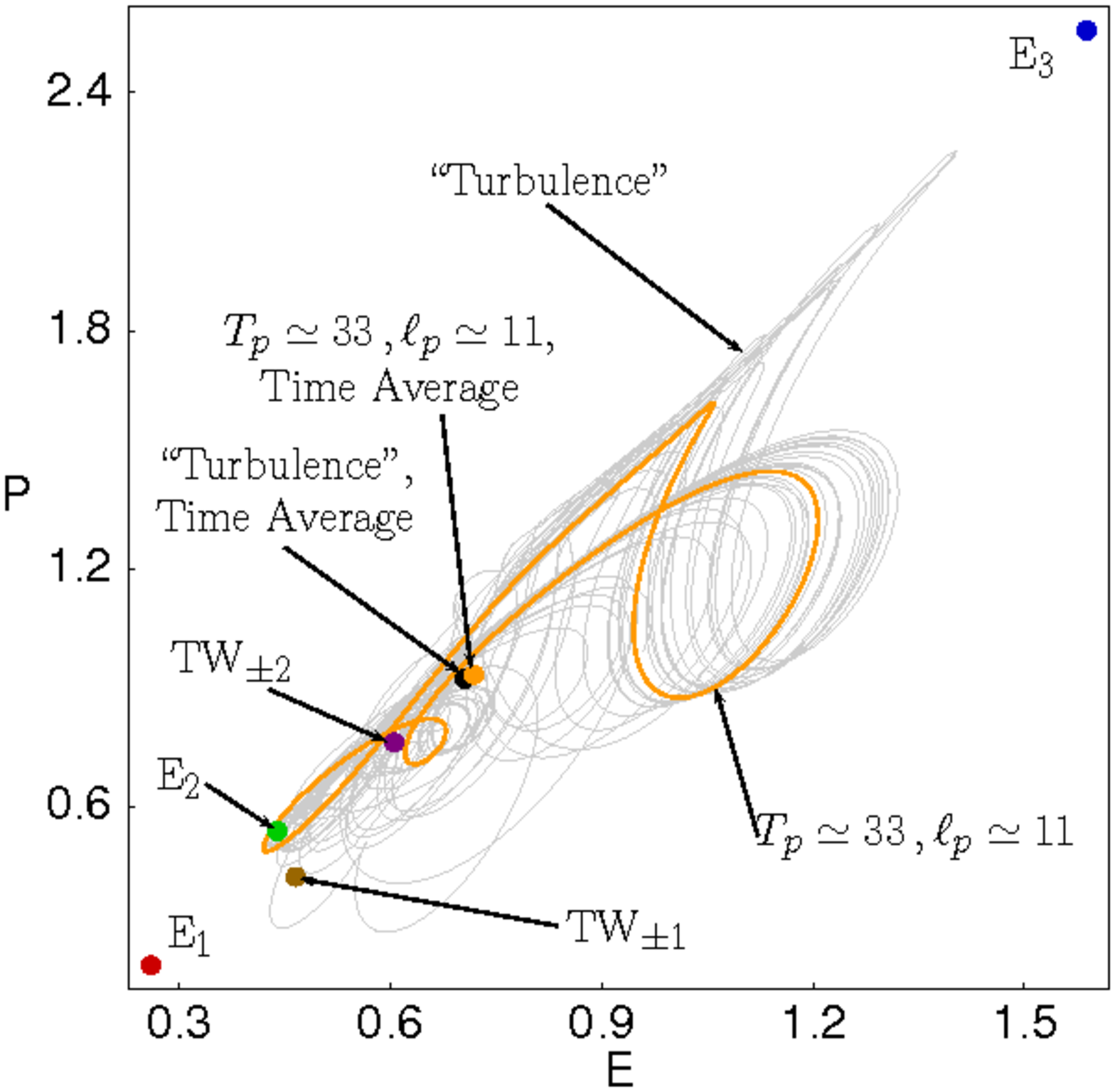}

  \end{tabular}
\end{center}
\caption{
(a) Power input $P$ {\em vs.}
dissipation rate $D$
(b) energy $E$  {\em vs.}
power input $P$,   for several  \eqva\ and \reqva,
a \rpo, and a typical `turbulent' long-time trajectory.
Projections of the heteroclinic connections are
given in \reffig{f:drivedragConn}.
System size $L=22$.
        }
\label{f:drivedrag}
\end{figure}
%%%%%%%%%%%%%%%%%%%%%%%%%%%%%%%%%%%%%%%%%%%%%%%%%%%%%%%%%%%%%%%%%%

%%%%%%%%%%%%%%%%%%%%%%%%%%%%%%%%%%%%%%%%%%%%%%%%%%%%%%%%%%%%%%%%
\begin{figure}[t]
\begin{center}
 \begin{tabular}{cc}
        ~~~~~~~~(\textit{a})                        &   ~~~~~~~~(\textit{b}) \\
    \includegraphics[width=0.46\textwidth, clip=true]{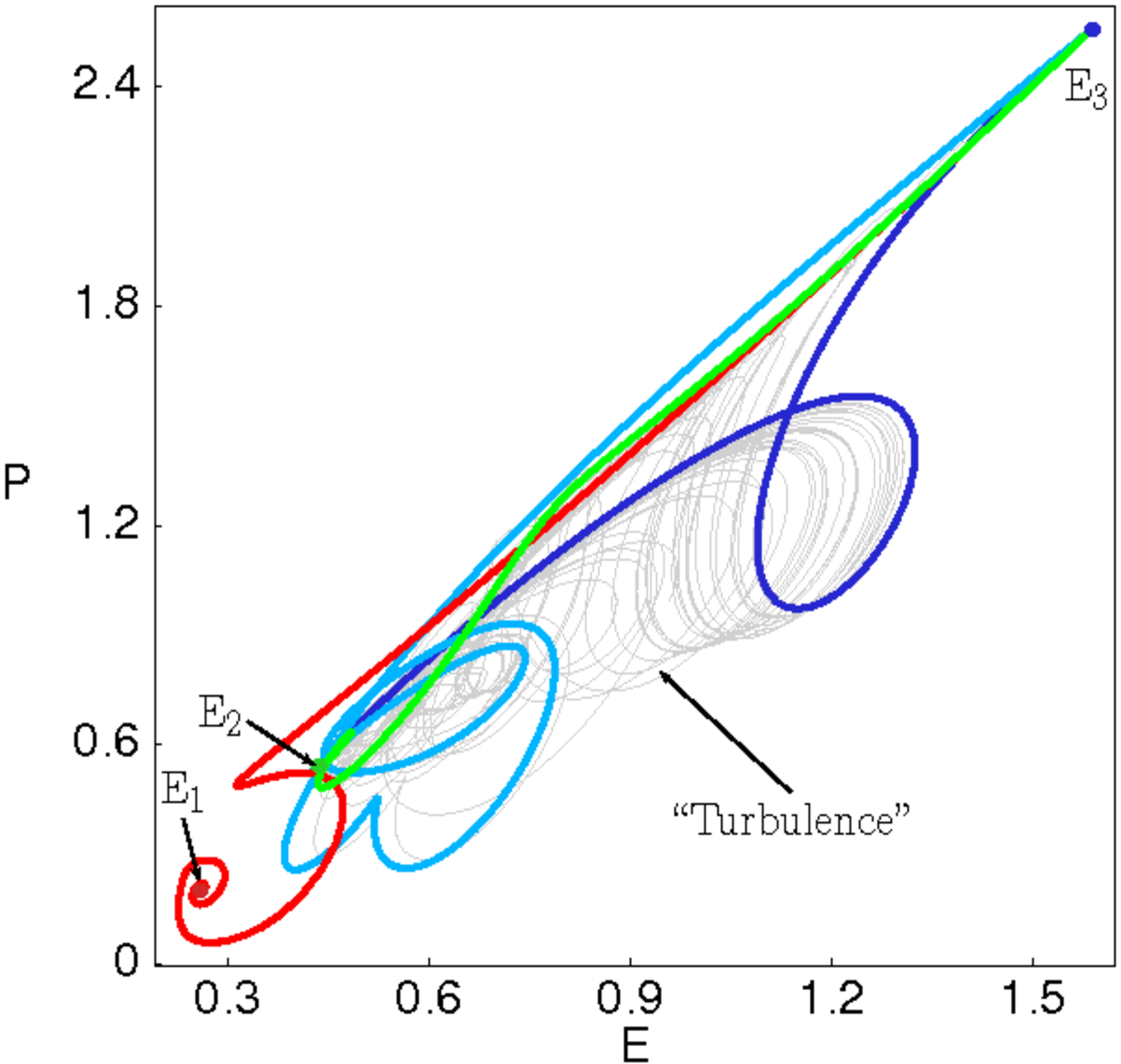}     & \includegraphics[width=0.46\textwidth, clip=true]{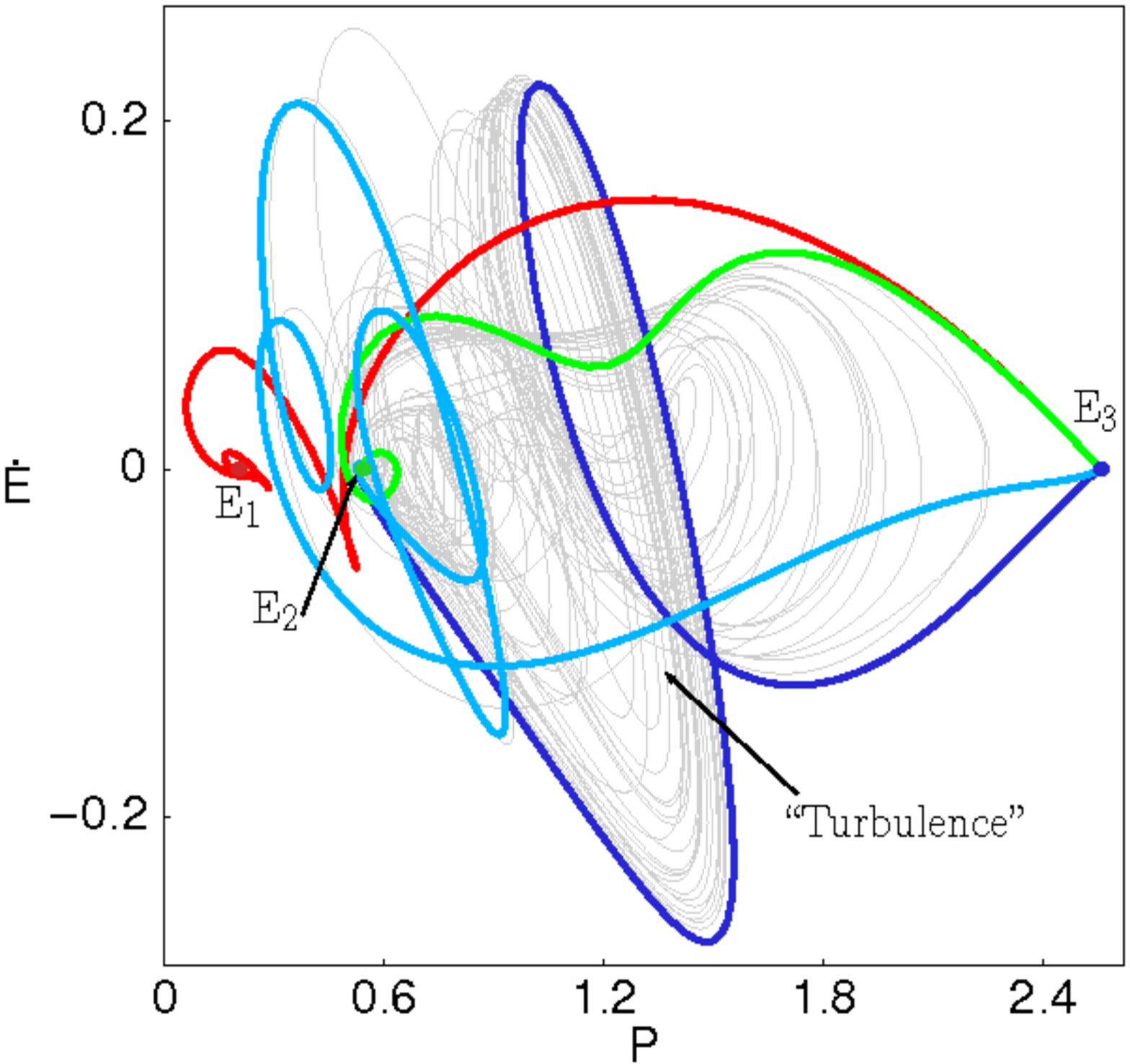}
 \end{tabular}
\end{center}
\caption{
Two projections of the $(E,P,\dot{E})$ representation of the flow.
\EQV{1} (red), \EQV{2} (green), \EQV{3} (blue),
heteroclinic connections from \EQV{2} to $\EQV{3}$ (green),
from $\EQV{1}$ to \EQV{3} (red)
and from \EQV{3} to $\EQV{2}$ (shades of blue), superimposed over
a generic long-time `turbulent' trajectory (grey).
(a) As in \reffig{f:drivedragConn}\,(\textit{b}),
with labels omitted for clarity.
(b) A plot of  $\dot{\expctE} = P - D$ yields a clearer
visualization than \reffig{f:drivedragConn}\,(\textit{a}).
System size $L=22$.
        }
\label{f:drivedragConn}
\end{figure}
%%%%%%%%%%%%%%%%%%%%%%%%%%%%%%%%%%%%%%%%%%%%%%%%%%%%%%%%%%%%%%%%%%

In \reffig{f:drivedrag} we plot \refeq{EnRate}, the time-dependent
$\dot{\expctE}$ in the power input $P$ {\em vs.}
dissipation rate $D$ plane, for $L=22$ \eqva\ and \reqva,
a selected \rpo, and for a typical `turbulent' long-time
trajectory.

Projections from the $\infty$-dimensional \statesp\ onto the 3-dimensional
$(E,P,D)$ representation of the flow, such as
\reffigs{f:drivedrag}{f:drivedragConn}, can be misleading.
The most one can say is that if points are clearly separated in an
$(E,P,D)$ plot (for example, in \reffig{f:drivedrag}
$\EQV{1}$ \eqv\ is outside the recurrent set), they are also separated
in the full \statesp.  Converse is not true -- states of
very different topology can have similar energies.

An example is
the \rpo\ $(\period{p},\shift_p) = (32.8,10.96)$
(see \reffig{f:ks22rpos}\,(\textit{b})) which is the least
unstable short \rpo\ we have detected in this system. It appears to be
well embedded within the turbulent flow. The mean power $\timeAver{P_p}$ evaluated
as in \refeq{poE}, see \reffig{f:drivedrag},
is numerically quite close to the long-time
turbulent time average $\timeAver{P}$.
Similarly close prediction of mean dissipation rate in the
\pCf\ from a single-period \po\ computed by
Kawahara and Kida\rf{KawKida01} has lead to
optimistic hopes that `turbulence' is different from
low-dimensional chaos, insofar that the determination of one special
\po\ could yield all long-time averages.
Regrettably, not true -- as always, here too one needs a hierarchy
of \po s of increasing length to obtain accurate
predictions\rf{DasBuch}.

%%%%%%%%%%%%%%%%%%%%%%%%%%%%%%%%%%%%%%%%%%%%%%%%%%%%%%%%%%%%%%%%
\begin{figure}[t]
\begin{center}
(\textit{a})\includegraphics[width=0.46\textwidth, clip=true]{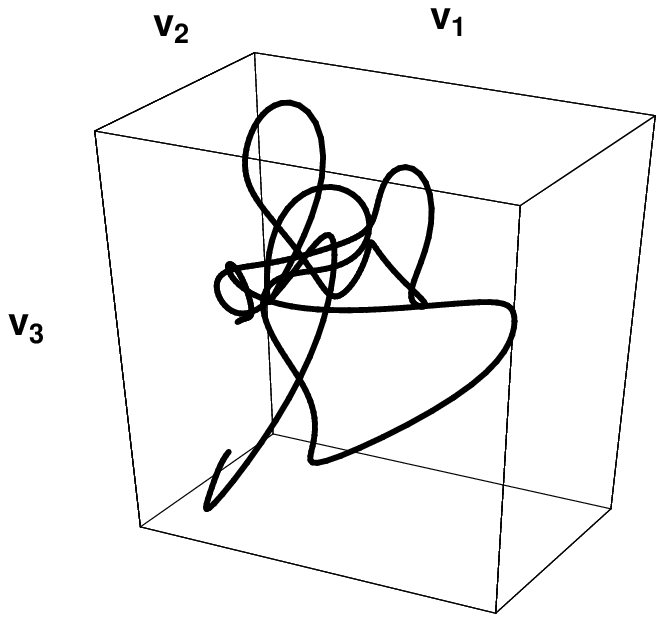}
(\textit{b})\includegraphics[width=0.46\textwidth, clip=true]{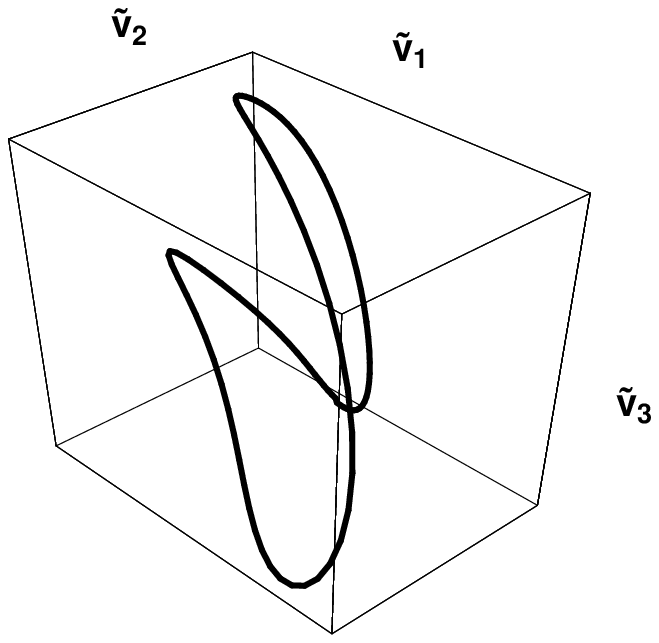}
\\
\end{center}
\caption{
 The
\rpo\ with $(\period{p},\shift_p) =(33.5,4.04)$
from \reffig{f:ks22rpos}\,(\textit{c})
which appears well embedded within the turbulent flow:
 (a) A stationary \statesp\ projection,
  traced for four periods $\period{p}$. The coordinate axes
$v_1$, $v_2$, and $v_3$ are those of \reffig{f:KS22E2man};
 (b) In the co-moving mean velocity frame.
        } \label{f:MeanVelocityFrame}
\end{figure}
%%%%%%%%%%%%%%%%%%%%%%%%%%%%%%%%%%%%%%%%%%%%%%%%%%%%%%%%%%%%%%%%%%

For any given \rpo\ a convenient visualization is
offered by the {\em mean velocity frame}, {\ie},
a reference frame that rotates with velocity
$\velRel_p=\shift_p/\period{p}$.
In the mean velocity frame a \rpo\ becomes
a \po, as in \reffig{f:MeanVelocityFrame}\,(\textit{b}).
However, each {\rpo} has its own mean velocity frame and thus
sets of \rpo s are difficult to visualize simultaneously.

%% file: summary.tex
% summary.tex
% $Author: siminos $ $Date: 2009-10-04 22:20:49 +0300 (Sun, 04 Oct 2009) $

\section{Summary}
\label{sect:rpo-sum}

In this paper we study the \KS\ flow as a staging ground for
testing dynamical systems approaches to
moderate Reynolds number turbulence in full-fledged
({\em not} a few-modes model),
infinite-dimensional \statesp\ PDE settings\rf{Holmes96},
and present a detailed geometrical portrait of dynamics in the
{\KS} \statesp\ for the $L=22$ system size, the smallest
system size for which this system empirically exhibits
`sustained turbulence.'

Compared to the earlier work
\rf{Christiansen97,LanThesis,lanCvit07,lop05rel},
the main advances here are the new insights in
the role that continuous symmetries,
discrete symmetries,
low-dimensional unstable manifolds of \eqva,
and the connections between \eqva\ play in organizing the flow.
The key new feature of the translationally invariant KS
on a periodic domain
are the attendant continuous families of
\reqva\ (traveling waves) and \rpo s.
We have now understood the preponderance of solutions of
relative type, and lost fear of them:
a large number of unstable \rpo s and \po s has been determined
here numerically.

Visualization of infinite-dimensional
\statesp\ flows, especially in presence of continuous symmetries,
is not straightforward.
At first glance, turbulent dynamics visualized in the \statesp\ appears
hopelessly complex, but under a detailed examination it is
much less so than feared: for strongly dissipative flows (KS, Navier-Stokes)
it is pieced together from low dimensional
local unstable manifolds connected by fast transient interludes.
In this paper we offer two low-dimensional visualizations of such
flows: (1) projections onto 2- or 3-dimensional,
PDE representation independent
dynamically invariant frames, and
(2) projections onto
the physical, symmetry invariant but time-dependent
energy transfer rates.

\Rpo s require a reformulation of the periodic orbit
theory\rf{Cvi07}, as well as a rethinking of the dynamical
systems approaches to constructing symbolic dynamics,
outstanding problems that we hope to address in near future%
\rf{SCD09b,SiminosThesis}.
What we have learned from the $L=22$ system  is that many of
these \rpo s appear organized by the unstable manifold of
$\EQV{2}$, closely following the homoclinic loop formed
between $\EQV{2}$ and $\Shift_{1/4}\EQV{2}$.

In the spirit of the parallel studies of boundary shear flows\rf{HaKiWa95},
the {\KS} $L=22$ system size was chosen as the smallest
system size for which KS empirically exhibits
`sustained turbulence.'
This is convenient both for
the analysis of the \statesp\ geometry, and for the numerical reasons,
 but the price is high - much of the
observed dynamics is specific to this unphysical, externally
imposed periodicity. What needs to be
understood is the nature of \eqv\ and \rpo\ solutions in the
$L \to \infty$ limit, and the structure of the $L = \infty$ periodic orbit
theory.

In summary, {\KS} (and \pCf, see \refref{GHCW07})  \eqva, \reqva, \po s and
\rpo s embody Hopf's vision\rf{hopf48}: together they form the
 repertoire of recurrent spatio-temporal
patterns explored by turbulent dynamics.

%% file: ackn.tex
% ackn.tex
% $Author: siminos $ $Date: 2009-10-04 22:20:49 +0300 (Sun, 04 Oct 2009) $

\section*{Acknowledgments}

We are grateful to Y.~Lan for pointing out to us the existence of
the  \EQV{1}~\eqv\ at the $L=22$ system size, J.~Crofts for a key
observation \rf{Crofts07thesis} that led to faster \rpo\ searches,
J.F.~Gibson for many spirited exchanges, and the anonymous referee
for many perspicacious observations.
P.C. and E.S. thank
G.~Robinson,~Jr.\ for support.
E.S. was partly supported by NSF grant DMS-0807574.
R.L.D. gratefully acknowledges the support from EPSRC under grant GR/S98986/01.

%% file: fourierRLD.tex
% fourierRLD.tex
% $Author: siminos $ $Date: 2009-10-04 22:20:49 +0300 (Sun, 04 Oct 2009) $

\section{Integrating \KSe\ numerically} \label{sec:fourierRLD}

The \KSe\ in terms of Fourier modes:
\beq
  \hat{u}_k = {\cal F}[u]_k = \frac{1}{L}\int_0^L u(x,t) e^{-i q_kx}dx\,,
  \qquad u(x,t) = {\cal F}^{-1}[\hat{u}] =
  \sum_{k\in{\mathbb Z}} \hat{u}_k e^{i q_k x}
\eeq
is given by
\beq
  \dot{\hat{u}}_k = \left(q_k^2-q_k^4\right) \hat{u}_k -
  \frac{i q_k}{2}{\cal F}[({\cal F}^{-1}[\hat{u}])^2]_k\,.
\eeq
Since $u$ is real, the Fourier modes are related by $\hat{u}_{-k} =
\hat{u}^\ast_k$.

The above system is truncated as follows: The Fourier transform
${\cal F}$ is replaced by its discrete equivalent
\beq
  a_k = {\cal F}_N[u]_k = \sum_{n = 0}^{N-1} u(x_n)
  e^{-i q_k x_n}\,,\qquad u(x_n) = {\cal F}_N^{-1}[a]_n
  = \frac{1}{N}\sum_{k = 0}^{N-1} a_k e^{i q_k x_n}\,,
\eeq
where $x_n = nL/N$ and $a_{N-k} = a^\ast_k$.  Since $a_0
= 0$ due to Galilean invariance and setting $a_{N/2} = 0$ (assuming
$N$ is even), the number of independent variables in the truncated
system is $N-2$:
\beq
  \dot{a}_k = \pVeloc_k(a) = \left(q_k^2-q_k^4\right)a_k -
  \frac{i q_k}{2}{\cal F}_N[({\cal F}_N^{-1}[a])^2]_k\,,
\ee{eq:KS}
where $k = 1,\ldots,N/2-1$, although in the Fourier transform we need
to use $a_k$ over the full range of $k$ values from 0 to $N-1$.
As $a_k \in \mathbb{C}$, \refeq{eq:KS} represents a
system of ordinary differential equations in ${\mathbb R}^{N-2}$.

The discrete Fourier transform ${\cal F}_N$ can be computed by FFT.
In Fortran and C, the FFTW library \refref{Frigo05} can be used.

In order to find the \jacobianM\ of the solution, or compute Lyapunov
exponents of the \KS\ flow, one needs to solve the equation for a
displacement vector $b$ in the tangent space: \beq
  \dot{b} = \frac{\partial \pVeloc(a)}{\partial a} b\,.
\eeq
Since ${\cal F}_N$ is a linear operator, it is easy to show that
\beq
  \dot{b_k} = \left(q_k^2-q_k^4\right)b_k -
  iq_k{\cal F}_N[{\cal F}_N^{-1}[a]\otimes {\cal F}_N^{-1}[b]]_k\,,
\ee{eq:KStan}
where $\otimes$ indicates componentwise product of two
vectors, \ie, $a\otimes b = \diag(a)\,b = \diag(b)\,a$.  This equation
needs to be solved simultaneously with \refeq{eq:KS}.

Equations \refeq{eq:KS} and \refeq{eq:KStan} were solved using the
exponential time differencing 4th-order Runge-Kutta method
(ETDRK4)\rf{cox02jcomp,ks05com}.

\section{Determining stability properties of equilibria,
traveling waves, and \rpo s} \label{sec:stability}

Let $f^t$ be the flow map of the \KSe, \ie\ $f^t(a) = a(t)$ is the
solution of \refeq{eq:KS} with initial condition $a(0) = a$.
The stability properties of the solution $f^t(a)$ are
determined by the fundamental matrix $J(a,t)$ consisting of partial
derivatives of $f^t(a)$ with respect to $a$.  Since $a$ and
$f^t$ are complex valued vectors, the real valued matrix
$J(a,t)$ contains partial derivatives evaluated separately with
respect to the real and imaginary parts of $a$, that is
\beq
  J(a,t) = \frac{\partial f^t(a)}{\partial a}
  = \left(\begin{array}{cccc}
  \frac{\partial f^t_{R,1}}{\partial a_{R,1}} &
  \frac{\partial f^t_{R,1}}{\partial a_{I,1}} &
  \frac{\partial f^t_{R,1}}{\partial a_{R,2}} & \\[1ex]
  \frac{\partial f^t_{I,1}}{\partial a_{R,1}} &
  \frac{\partial f^t_{I,1}}{\partial a_{I,1}} &
  \frac{\partial f^t_{I,1}}{\partial a_{R,2}} & \cdots \\[1ex]
  \frac{\partial f^t_{R,2}}{\partial a_{R,1}} &
  \frac{\partial f^t_{R,2}}{\partial a_{I,1}} &
  \frac{\partial f^t_{R,2}}{\partial a_{R,2}} & \\
  & \vdots & & \ddots \end{array}\right)
\label{eq:FundMat}\eeq
where $a_k = a_{R,k} + i a_{I,k}$ and $f^t_k = f^t_{R,k} + i f^t_{I,k}$.
The partial derivatives $\frac{\partial f^t}{\partial a_{R,j}}$
and $\frac{\partial f^t}{\partial a_{I,j}}$ are determined
by solving \refeq{eq:KStan} with initial conditions
$b_k(0) = b_{N-k}(0) = 1 + 0i$ and $b_k(0) = -b_{N-k}(0) = 0 + 1i$,
respectively, for $k = j$ and $b_k(0) = 0$ otherwise.

The stability of a \po\ with period $\period{p}$ is determined by the location
of eivenvalues of $J(a_p,\period{p})$ with respect to the unit circle in the
complex plane.

Because of the translation invariance, the stability of a \rpo\ is
determined by the eigenvalues of the matrix
$\mathbf{g}(\shift_p)\,J(a_p,\period{p})$, where $\mathbf{g}(\shift)$
is the action of the translation operator introduced in
\refeq{eq:shiftFour}, which in real valued representation takes the form
of a block diagonal matrix with the $2\times 2$ blocks
\[
  \left( \begin{array}{cc}
   \cos q_k\shift  & \sin q_k\shift \\
   -\sin q_k\shift & \cos q_k\shift
  \end{array}\right),\ \ k=1,2,\ldots, N/2-1\,.
\]

For an equilibrium solution $a_q$, $f^t(a_q) = a_q$ and so
the fundamental matrix $J(a_q,t)$ can be expressed in terms of the
time independent stability matrix $A(a_q)$ as follows
\[  J(a_q,t) = e^{A(a_q)t}, \]
where
\beq
  A(a_q) = \left.\frac{\partial v}{\partial a}\right|_{a=a_q}.
\label{eq:StabMat}\eeq
Using the real valued representation of \refeq{eq:FundMat},
the partial derivatives of $v(a)$ with respect to the real and imaginary
parts of $a$ are given by
\bea
    \frac{\partial v_k}{\partial a_{R,j}} & = &
    \left(q_k^2 - q_k^4\right)\delta_{kj}
    - iq_k {\cal F}_N[{\cal F}_N^{-1}[a]\otimes {\cal F}_N^{-1}[b_R^{(j)}]]_k\,,
\continue
    \frac{\partial v_k}{\partial a_{I,j}} & = &
    \left(q_k^2 - q_k^4\right)i\delta_{kj}
    - iq_k {\cal F}_N[{\cal F}_N^{-1}[a]\otimes {\cal F}_N^{-1}[b_I^{(j)}]]_k\,,
\label{eq:StabMatElem}
\eea
where $b_R^{(j)}$ and $b_I^{(j)}$ are complex valued vectors such that
$b_{R,k}^{(j)} = b_{R,N-k}^{(j)} = 1 + 0i$ and
$b_{I,k}^{(j)} = -b_{I,N-k}^{(j)} = 0 + 1i$ for $k = j$ and
$b_{R,k}^{(j)} = b_{I,k}^{(j)} = 0$ otherwise.
In terms of $a_{R,k}$ and $a_{I,k}$ we have
\bea
    \frac{\partial v_{R,k}}{\partial a_{R,j}} & = &
    \left(q_k^2 - q_k^4\right)\delta_{kj}
    + q_k (a_{I,k+j} + a_{I,k-j})\,,
\continue
    \frac{\partial v_{R,k}}{\partial a_{I,j}} & = &
    - q_k (a_{R,k+j} - a_{R,k-j})\,,
\label{expanMvar}\\
    \frac{\partial v_{I,k}}{\partial a_{R,j}} & = &
    - q_k (a_{R,k+j} + a_{R,k-j})\,,
\continue
    \frac{\partial v_{I,k}}{\partial a_{I,j}} & = &
    \left(q_k^2 - q_k^4\right)\delta_{kj}
    - q_k (a_{I,k+j} - a_{I,k-j})\,,
\nnu
\eea
where $\delta_{kj}$ is Kronecker delta.

The stability of equilibria is characterized by the sign
of the real part of the eigenvalues of $A(a_q)$.
The stability of a \reqv\ is detemined in the co-moving reference frame, so
the fundamental matrix takes the form $\mathbf{g}(\velRel_q t)\,J(a_q,t)$.  The
stability matrix of a \reqv\ is thus equal to $A(a_q) + \velRel_q \translGen$
where $\translGen = iq_k\delta_{kj}$ is the Lie algebra translation
generator, which in the real-space representation takes the form
$ \translGen = \diag(0, q_1, 0, q_2, \ldots)$.

%% file: lmderRLD.tex
% lmderRLD.tex
% $Author: siminos $ $Date: 2009-10-04 22:20:49 +0300 (Sun, 04 Oct 2009) $

\section{Levenberg--Marquardt searches for \rpo s}
\label{sec:lmderRLD}

To find \rpo s of the \KS\ flow, we use multiple shooting and
the Levenberg--Marquardt (LM) algorithm implemented in the routine
{\tt lmder} from the MINPACK software package\rf{minpack}.

In order to find periodic orbits, a system of nonlinear
algebraic equations needs to be solved.  For flows, this
system is underdetermined, so, traditionally, it is augmented
with a constraint that restricts the search space to be
transversal to the flow (otherwise, most of the popular
solvers of systems of nonlinear algebraic equations, e.g.
those based on Newton's method, cannot be used). When
detecting \rpo s, a constraint is added for each continuous
symmetry of the flow.  For example, when detecting relative
periodic orbits in the complex Ginzburg Landau equation,
L{\'o}pez {\etal}\rf{lop05rel} introduce three additional
constraints.

Our approach differs from those used previously in that we do
not introduce the constraints.  Being an optimization solver,
the LM algorithm has no problem with solving an
underdetermined system of equations, and, even though {\tt
lmder} explicitly restricts the number of equations to be not
smaller than the number of variables, the additional
equations can be set identically equal to
zero\rf{Crofts07thesis}.  In fact, there is numerical
evidence that, when implemented with additional constraints,
the solver usually takes more steps to converge from the same
seed, or fails to converge at all\rf{Crofts07thesis}. In what
follows we give a detailed description of the algorithm and
the search strategy which we have used to find a large number
of \rpo s defined in \refeq{KSrpos} and pre-periodic orbits
defined in \refeq{KSpos}.

When searching for \rpo s of truncated \KS\ equation
\refeq{eq:KS}, we need to solve the system of $N-2$ equations
\beq
  {\bf g}(\shift)f^\period{}(a) - a = 0\,,
\ee{eq:system}
with $N$ unknowns $(a, \period{}, \shift)$, where $f^t$
is the flow map of the \KSe.  In the case of pre-periodic orbits, the system has
the form
\beq
  -{\bf g}(-\shift)[f^\period{}(a)]^\ast - a = 0\,,
\ee{eq:systemppos}
(see \refeq{KSposFour}).

We have tried two different implementations of the multiple shooting.
The emphasis was on the simplicity of the implementations, so, even
though both implementations worked equally well, each of them had
its own minor drawbacks.

In the first implementation, we fix the total number of steps within
each shooting stage and change the numerical integrator step size
$h$ in order to adjust the total integration time to a desired value
$\period{}$.

Let $(\hat{a}, \hat{\period{}}, \hat{\shift})$ be the starting guess for a \rpo\
obtained through a close return within a chaotic attractor (see below).
We require that the initial integration step size
does not exceed $h_0$, so we round off the
number of integration steps to $n = \lceil \hat{\period{}}/h_0\rceil$, where
$\lceil x \rceil$ denotes the nearest integer larger than $x$.

The integration step size is equal to $h = \period{}/n$. With the
number of shooting stages equal to $m$, the system in
\refeq{eq:system} is rewritten as follows
\begin{eqnarray}\label{eq:MultShoot}
 F^{(1)}&\!=\!& f^\tau(a^{(1)}) - a^{(2)} = 0\,,\nonumber\\
 F^{(2)}&\!=\!& f^\tau(a^{(2)}) - a^{(3)} = 0\,,\nonumber\\
 && \cdots \\
 F^{(m-1)}&\!=\!& f^\tau(a^{(m-1)}) - a^{(m)} = 0\,,\nonumber\\
 F^{(m)}&\!=\!& {\bf g}(\shift)f^{\tau'}(a^{(m)}) - a^{(1)} = 0\,,\nonumber
\end{eqnarray}
where $\tau = \lfloor n/m \rfloor h$ ($\lfloor x \rfloor$ is the nearest
integer smaller than $x$),
$\tau' = nh - (m-1)\tau$, and $a^{(j)} = f^{(j-1)\tau}(a)$,
$j = 1, \ldots , m$.  For the detection of pre-periodic orbits, the last equation
in \refeq{eq:MultShoot} should be replaced with
\[
 F^{(m)} = -{\bf g}(-\shift)[f^{\tau'}(a^{(m)})]^\ast - a^{(1)} = 0\,.
\]
With the \jacobianM\ of \refeq{eq:MultShoot} written as
\beq
  J = \left(\begin{array}{ccc}\!\!
   \displaystyle \frac{\partial F^{(j)}}{\partial a^{(k)}} &
   \displaystyle \frac{\partial F^{(j)}}{\partial \period{}} &
   \displaystyle \frac{\partial F^{(j)}}{\partial \shift}\!\!
  \end{array}\right),\quad j,k = 1,\ldots,m\,,
\eeq
the partial derivatives with respect to $a^{(k)}$ can be calculated
using the solution of \refeq{eq:KStan} as described in
\refappe{sec:stability}.  The partial derivatives
with respect to $T$ are given by
\beq
  \frac{\partial F^{(j)}}{\partial \period{}} =
  \left\{\begin{array}{ll}
    \frac{\partial f^\tau(a^{(j)})}{\partial \tau}
    \frac{\partial \tau}{\partial T} = v(f^\tau(a^{(j)}))
    \lfloor n/m \rfloor/n\,, & j = 1,\ldots, m-1\\[.5ex]
    {\bf g}(\shift) v(f^{\tau'}(a^{(j)}))
    (1 - \frac{m-1}{n} \lfloor n/m \rfloor ), & j = m\,.
  \end{array}\right.
\eeq
Note that, even though $\partial f^t(a) /\partial t = v(f^t(a))$,
it should not be evaluated using the equation for the vector field $v$.
The reason is that, since the flow $f^t$ is approximated by a
numerical solution, the derivative of the numerical solution with
respect to the step size $h$ may differ from the vector field $v$,
especially for larger step sizes.  We evaluate the derivative by
a forward difference using numerical integration with step sizes
$h$ and $h + \delta$:
\beq
  \frac{\partial f^{jh}(a)}{\partial t} = \frac{1}{j\delta}
  \left[f^{j(h+\delta)}(a) - f^{jh}(a)\right],\quad j \in
  {\mathbb Z}^{+}
\eeq with $t = jh$ and $\delta = 10^{-7}$ for double precision
calculations. Partial derivatives $\partial F^{(j)}/\partial \shift$
are all equal to zero except for $j = m$, where it is given by
\beq
  \frac{\partial F^{(m)}}{\partial \shift} =
  \frac{d{\bf g}}{d\shift}f^{\tau'}(a^{(m)}) =
  \diag(i q_k e^{i q_k\, \shift} )f^{\tau'}(a^{(m)})\,.
\eeq

This \jacobianM\ is supplied to {\tt lmder}
augmented with two rows of zeros corresponding to the two identical
zeros augmenting \refeq{eq:MultShoot} in order to make the number of
equations formally equal to the number of variables,
as discussed above.

In the second implementation, we keep $h$ and $\tau$ fixed and vary
only $\tau' = \period{} - (m-1)\tau$.  In this case, we need to be
able to determine the numerical solution of \KSe\ not only at times
$t_j = jh, j = 1, 2, \ldots$, but at any intermediate time as well.
We do this by a cubic polynomial interpolation through points
$f^{t_j}(a)$ and $f^{t_{j+1}}(a)$ with slopes $v(f^{t_j}(a))$ and
$v(f^{t_{j+1}}(a))$.  The difference from the first implementation
is that partial derivatives $\partial F^{(j)}/\partial \period{}$
are zero for all $j = 1,\ldots,m-1$, except for
\beq
  \frac{\partial F^{(m)}}{\partial \period{}} =
  {\bf g}(\shift)v(f^{\tau'}(a^{(m)}))\,.
\eeq
which, for consistency, needs to be evaluated from the cubic
polynomial, not from the flow equation evaluated
at $f^{\tau'}(a^{(m)})$.

For detecting \rpo s of the \KS\ flow with $L = 22$, we used
$N = 32$, $h = 0.25$ (or $h_0 = 0.25$ within the first implementation),
and a number of shooting stages such that $\tau \approx 40.0$.
While both implementations were equally successful in detecting
periodic orbits of \KS\ flow, we found the second implementation more
convenient.

%%%%%%%%%%%%%%%%%%%%%%%%%%%%%%%%%%%%%%%%%%%%%%%%%%%%%%%%%%%%%%
\begin{figure}[t]
\begin{center}
\includegraphics[width=0.9\textwidth, clip=true]{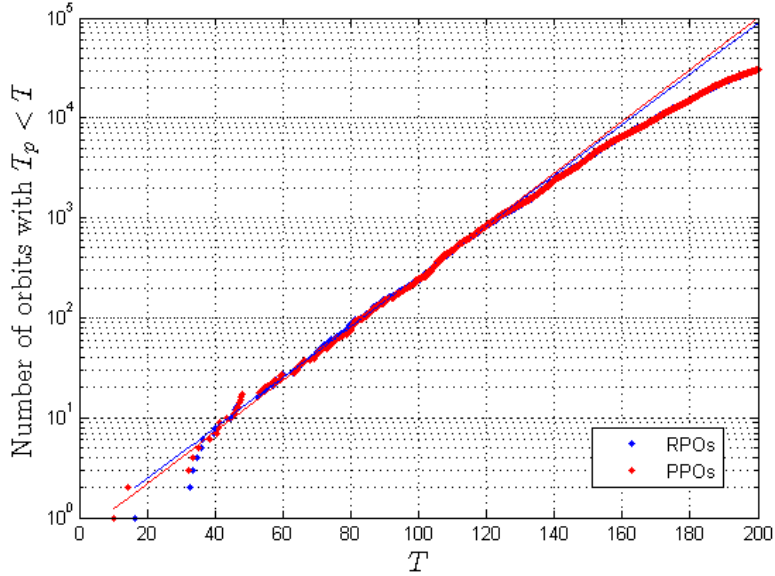}
\end{center}
\caption{
Numbers of detected \rpo s (RPOs) and pre-periodic orbits (PPOs)
with periods smaller than $T$.  The lines indicate the linear fit
to the logarithm of the number of orbits as functions of $T$ in the
range $T \in [70, 120]$.
     } \label{fig:Npos}
\end{figure}
%%%%%%%%%%%%%%%%%%%%%%%%%%%%%%%%%%%%%%%%%%%%%%%%%%%%%%%%%%%%%%%%%%

The following search strategy was adopted: The search for
\rpo s with $\period{} \in [10, 200]$ was conducted within a
rectangular region containing the chaotic attractor.  To
generate a seed, a random point was selected within the
region and the flow \refeq{eq:KS} was integrated for a
transient time $t = 40$, sufficient for an orbit to settle on
the attractor at some point $\hat{a}$.  This point was taken
to be the seed location.  In order to find orbits with
different periods, the time interval $[10, 200]$ was
subdivided into windows of length 10, i.e. $[t_\mathrm{min},
t_\mathrm{max}]$, where $t_\mathrm{min} = 10j$ and
$t_\mathrm{max} = 10(j+1)$, with $j = 1, 2, \ldots, 19$.  To
determine the seed time $\hat{\period{}}$ and shift
$\hat{\shift}$, we located an approximate global minimum of
$\| {\bf g}(\shift)f^t(a) - a \|$ (or of $\| -{\bf
g}(-\shift)[f^t(a)]^\ast - a \|$ in the case of pre-periodic
orbits) as a function of $t \in [t_\mathrm{min},
t_\mathrm{max}]$ and $\shift \in (-L/2, L/2]$.  We did this
simply by finding the minimum value of the function on a grid
of points with resolution $h$ in time and $L/50$ in $\shift$.

Approximately equal numbers of seeds were generated for the
detection of \rpo s and pre-periodic orbits and within each
time window.  The hit rate, i.e. the fraction of seeds that
converged to \rpo s or pre-periodic orbits, varied from about
70\% for windows with $t_\mathrm{max} \leq 80$ to about 30\%
for windows with $t_\mathrm{min} \geq 160$.  The total number
of hits for \rpo s and pre-periodic orbits was over $10^6$
each.  Each newly found orbit was compared, after factoring
out the translation and reflection symmetries, to those already
detected.  As the search progressed, we found fewer and fewer
new orbits, with the numbers first saturating for smaller
period orbits.  At the end of the search we could find very
few new orbits with periods $T < 120$.  Thus we found over
30\,000 distinct prime \rpo s with $\shift > 0$ and over
30\,000 distinct prime pre-periodic orbits with $T < 200$.

In \reffig{fig:Npos} we show the numbers of detected \rpo s
and pre-periodic orbits with periods less than $T$.  It shows
that the numbers of \rpo s and pre-periodic orbits are
approx. equal and that they grow exponentially with
increasing $T$ up to $T \sim 130$, so that we are mostly
missing orbits with $T > 130$.  The straight line fits to the
logarithm of the numbers of orbits in the interval $T \in
[70, 120]$, represented by the lines in \reffig{fig:Npos},
indicate that the total numbers of \rpo s and pre-periodic
orbits with $T < 200$ could be over $10^5$ each.

To test the structural stability of the detected orbits and
their relevance to the full Kuramoto-Sivashinsky PDE, the
numerical accuracy was improved by increasing the number of
Fourier modes ($N = 64$) and reducing the step size ($h =
0.1$). Only a handful of orbits failed this higher-resolution
test. These orbits were not included in the list of the
60,000+ orbits detected.